\documentclass[aps,prd,twocolumn,superscriptaddress,floatfix,noeprint]{revtex4-1}
\usepackage[utf8]{inputenc}

\usepackage{color}
\usepackage{siunitx}
\usepackage{graphicx}
\usepackage{float}
\usepackage{multirow}
\usepackage{import}
\usepackage{hyperref}
\usepackage[caption=false]{subfig}
\usepackage{url}
\hypersetup{
    colorlinks=true,
    linkcolor=[rgb]{0.183, 0.295, 0.633},
    citecolor=[rgb]{0.211, 0.332, 0.183},
    filecolor=magenta,
    urlcolor=[rgb]{0.501, 0.095, 0.095},
}

\DeclareGraphicsExtensions{.pdf,.PDF,.png,.PNG,.jpg,.mps,.jpeg,.jbig2,.jb2,.JPG,.JPEG,.JBIG2,.JB2}

\definecolor{spring}{rgb}{0.7,0.9,0.7}
\definecolor{brick}{rgb}{0.7,0.2,0.1}
\definecolor{redHL}{rgb}{1.0,0.5,0.5}

\def\rtHz{\ensuremath{\sqrt{\textrm{Hz}}}}
\def\Msol{\ensuremath{M_\odot}}

\newcommand{\tem}[1]{TEM$_{#1}$}

\newcommand{\llorange}{\SI{134}{Mpc}} \newcommand{\lhorange}{\SI{111}{Mpc}}

\newcommand{\NumEvents}{56} 

\newcommand{\sqzfreq}{\SI{50}{Hz}} \newcommand{\qrpnfreq}{\SI{50}{Hz}} 

\newcommand{\lhoXarmPower}{{194 $\pm$ 2}} \newcommand{\lhoYarmPower}{{207 $\pm$ 2}} \newcommand{\lloXarmPower}{{232 $\pm$ 15}} \newcommand{\lloYarmPower}{{245 $\pm$ 5}} \newcommand{\lhoAveArmPower}{{\SI{201}{kW}}} \newcommand{\lloAveArmPower}{{\SI{239}{kW}}} 

\newcommand{\llodutySummary}{$77.0\%$} \newcommand{\lhodutySummary}{$74.6\%$} \newcommand{\coincident}{$62.2\%$} 

\newcommand{\lhoOthreeaduty}{\num{71.2}} \newcommand{\lloOthreeaduty}{\num{75.7}} \newcommand{\lhoOthreebduty}{\num{78.8}} \newcommand{\lloOthreebduty}{\num{78.6}}

\newcommand{\lloOoneduty}{\num{55.3}} \newcommand{\lhoOoneduty}{\num{62.6}}  

\newcommand{\lloOtwoduty}{\num{65.8}} \newcommand{\lhoOtwoduty}{\num{70.6}}

\newcommand{\typicalSQZdBLHO}{2.0}

\newcommand{\typicalSQZBNSrangeincreaseLHO}{12\%}

\newcommand{\typicalSQZdBwithouterr}{2.7}

\newcommand{\typicalSQZBNSrangeincrease}{14\%}

\newcommand{\sqzangdetuning}{7}

\begin{document}

\title[]{Sensitivity and Performance of the Advanced LIGO Detectors in the Third Observing Run}
\author{A.~Buikema}
\affiliation{LIGO, Massachusetts Institute of Technology, Cambridge, MA 02139, USA}
\affiliation{LIGO Livingston Observatory, Livingston, LA 70754, USA}
\author{C.~Cahillane}
\affiliation{LIGO, California Institute of Technology, Pasadena, CA 91125, USA}
\affiliation{LIGO Hanford Observatory, Richland, WA 99352, USA}
\author{G.~L.~Mansell}
\affiliation{LIGO Hanford Observatory, Richland, WA 99352, USA}
\affiliation{LIGO, Massachusetts Institute of Technology, Cambridge, MA 02139, USA}
\author{C.~D.~Blair}
\affiliation{OzGrav, University of Western Australia, Crawley, Western Australia 6009, Australia}
\affiliation{LIGO Livingston Observatory, Livingston, LA 70754, USA}
\author{R.~Abbott}
\affiliation{LIGO, California Institute of Technology, Pasadena, CA 91125, USA}
\author{C.~Adams}
\affiliation{LIGO Livingston Observatory, Livingston, LA 70754, USA}
\author{R.~X.~Adhikari}
\affiliation{LIGO, California Institute of Technology, Pasadena, CA 91125, USA}
\author{A.~Ananyeva}
\affiliation{LIGO, California Institute of Technology, Pasadena, CA 91125, USA}
\author{S.~Appert}
\affiliation{LIGO, California Institute of Technology, Pasadena, CA 91125, USA}
\author{K.~Arai}
\affiliation{LIGO, California Institute of Technology, Pasadena, CA 91125, USA}
% ...
 
\date{\today}

\begin{abstract}
On April 1st, 2019, the Advanced Laser Interferometer Gravitational-Wave Observatory (aLIGO), joined by the Advanced Virgo detector, began the third observing run, a year-long dedicated search for gravitational radiation.
The LIGO detectors have achieved a higher duty cycle and greater sensitivity to gravitational waves than ever before, with LIGO Hanford achieving angle-averaged sensitivity to binary neutron star coalescences to a distance of \lhorange, and LIGO Livingston to \llorange\ with duty factors of \lhodutySummary\ and \llodutySummary\ respectively.
The improvement in sensitivity and stability is a result of several upgrades to the detectors, including
doubled intracavity power,
the addition of an in-vacuum optical parametric oscillator for squeezed-light injection,
replacement of core optics and end reaction masses,
and installation of acoustic mode dampers.
This paper explores the purposes behind these upgrades, and explains to the best of our knowledge the noise currently limiting the sensitivity of each detector.
\end{abstract}

\maketitle

\tableofcontents 

\section{Introduction}\label{s:introduction}
In 2015, the Advanced LIGO detectors at Hanford, Washington and Livingston, Louisiana achieved unprecedented sensitivity to gravitational waves \cite{PhysRevD.93.112004,PhysRevLett.116.131103}. On September 14th, 2015, LIGO first detected gravitational waves from a binary black hole merger \cite{GW150914}.
During the first observing run (O1), which ran from September 2015 to January 2016, two more binary black hole detections were made~\cite{GW151226, O1:BBH}.
The second observing run (O2), which ran from November 2016 to August 2017, detected seven binary black hole mergers, and one binary neutron star merger~\cite{O2Catalog, GW170814, GW170608, GW170104, GW170817, GW170817_paper, PhysRevX.9.011001}.
The third observing run (O3), which ran from April 1 to September 30, 2019 (O3a) and from November 1, 2019 until March 27, 2020 (O3b), has been the most successful search for gravitational waves in history, with greater sensitivity and the permanent addition of the Advanced Virgo detector \cite{Abbott2018}.
During this run, \NumEvents{} candidate gravitational-wave signals, including at least one new compact binary coalescence in the binary neutron star mass range \cite{Abbott:2020uma} and a system with record mass ratio \cite{Abbott_2020}, were announced \cite{gracedbO3}.
The increase in the detection rate is due to the improved performance of the detectors, which is the subject of this paper.

\begin{figure}[hbt!]
  \centering
  \includegraphics[width=\columnwidth]{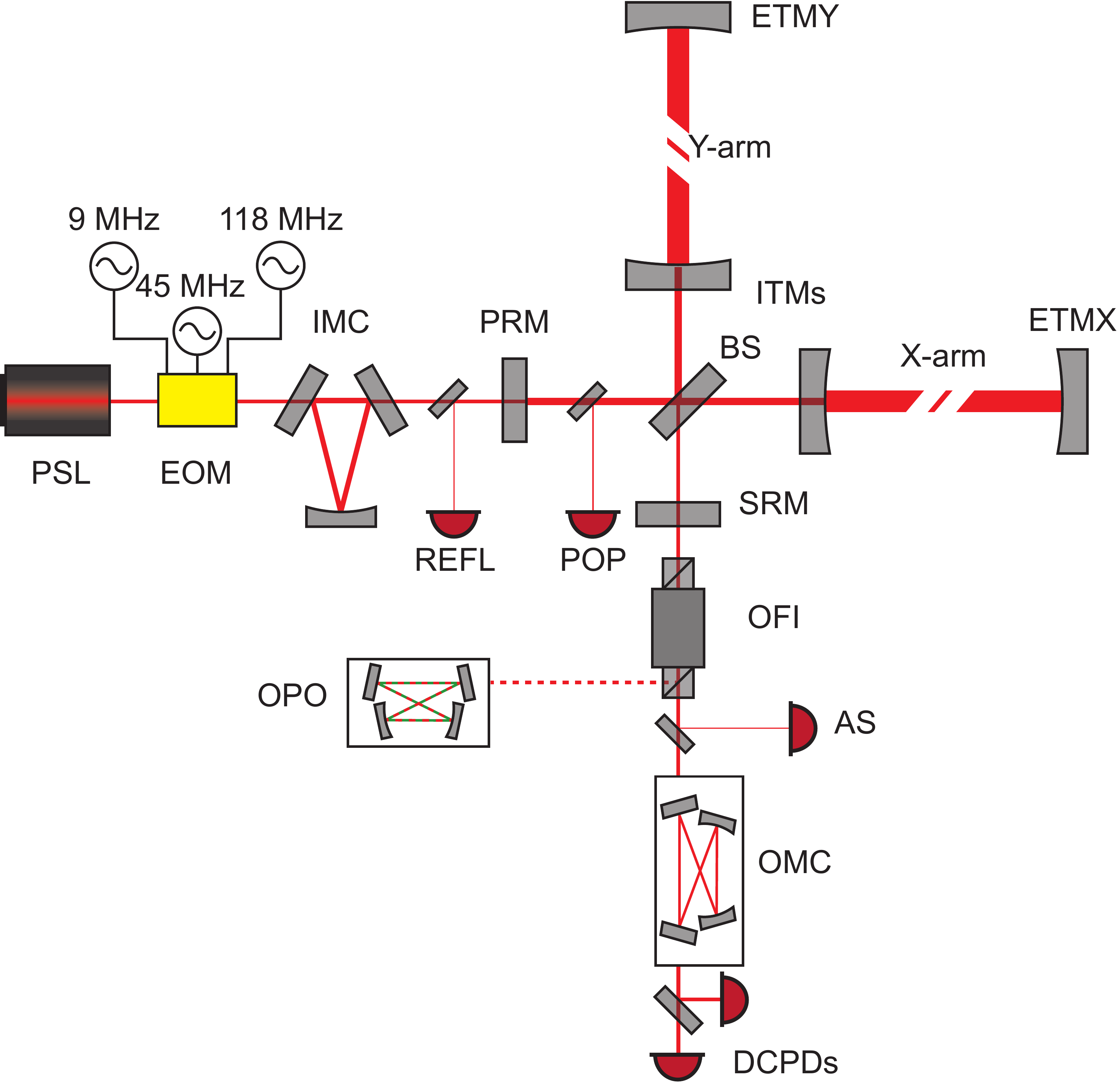}
  \caption{
  Simplified optical layout of the aLIGO detectors for O3.
  At the input port is the pre-stabilized laser (PSL) and the phase modulating electro-optic modulator (EOM) with three radio frequencies used for length and angular control.
  The spatial profile, polarization, jitter, and frequency noise of the beam is cleaned by the triangular input mode cleaner (IMC) cavity.
  Arm cavities are formed from input test masses (ITMs) and end test masses (ETMs).
  The power- and signal-recycling mirrors (PRM and SRM, respectively) together with the beamsplitter (BS) and input test masses form the power- and signal-recycling cavities.
  Light at the pick-off of the power-recycling cavity (POP) and interferometer reflection (REFL) ports are used for sensing and control of auxiliary degrees of freedom.
  The output Faraday isolator (OFI) prevents back-reflected light from entering the interferometer from the antisymmetric port (AS) and is used to inject squeezed light from the optical parametric oscillator (OPO).
  The output mode cleaner (OMC) cleans the output spatial mode and separates the carrier light for the output photodiodes (DCPDs) on transmission of the OMC.
  These photodiodes measure the differential arm length (DARM), which is sensitive to gravitational waves.
  }
  \label{fig:opticalLayout}
\end{figure}

The Advanced LIGO detectors are dual-recycled Fabry-P\'{e}rot Michelson interferometers.
Figure \ref{fig:opticalLayout} shows the full interferometer layout.
Ultra stable laser light at \SI{1064}{nm} \cite{Kwee2012, Thies2019} enters the interferometer and circulates in the \SI{4}{km} arms, each with Fabry-P\'{e}rot cavities to increase the light interaction time with a gravitational wave.
The power-recycling cavity, formed by the power-recycling mirror and the input test masses, increases the laser power circulating in the interferometer \cite{Meers1988}.
The signal-recycling cavity, formed with the signal-recycling mirror and the input test masses, broadens the detector bandwidth \cite{Mizuno_1993, BuonannoChen2001}.
Key parameters for both interferometers are summarized in Table \ref{table:aLIGOParams}.
Improvements made between observing runs bring the detectors closer to the final design sensitivity \cite{advancedLIGO}.

Gravitational waves passing through the interferometer produce a metric disturbance that results in an effective differential change in the arm lengths.  The change in effective arm length imparts a phase shift to the electromagnetic fields circulating in the arms.
This causes a change in optical power at the antisymmetric port via the interference between the fields from the two arms.
The gravitational-wave readout is a measure of the differential arm length, or DARM.

The lengths of these key optical cavities, and other auxiliary optical cavities like the input and output mode cleaners, are controlled using the length sensing and control system \cite{AdvLIGOFinalDesign, Izumi2017}.
Most cavity lengths are sensed using radio frequency phase modulation sidebands, added to the main beam by the electro-optic modulator, via the Pound-Drever-Hall laser stabilization technique~\cite{PoundDreverHall}.
Exceptions are the output mode cleaner, which uses a dither scheme described in Section~\ref{ss:omcl}, and DARM, which uses a DC readout scheme \cite{Fricke2012}.
The beat between carrier and sidebands present at the various ports of the interferometer---on reflection of the power-recycling mirror, at the pick-off of the power-recycling cavity, or at the antisymmetric port---is measured on photodetectors, filtered through a combination of analog and digital electronics, and then fed back to the relevant actuators via the LIGO real-time digital control system \cite{bork2020advligorts}.

Calibration is the process of characterizing the response of the detector to gravitational waves.
The DARM control loop and interferometer sensitivity are referenced to a photon-calibrator, which induces a known displacement on an end test mass via radiation pressure \cite{doi:10.1063/1.4967303}.
The uncertainty in the detector response to gravitational waves is 7\% in magnitude and 4 degrees in phase between 20 and \SI{2000}{Hz} \cite{Sun2020, PhysRevD.96.102001, Viets_2018, Tuyenbayev_2016}.

The alignment of optics in the interferometer is controlled by the alignment sensing and control system \cite{Barsotti2010}.
Three separate techniques are used: radio frequency wavefront sensors \cite{Fritschel:1998}, beam pointing onto quadrant photodiodes, and dither alignment described in Section~\ref{ss:ASC}.
Controlled degrees of freedom include the alignment of the input mode cleaner, input beam pointing, power- and signal-recycling cavities, Michelson, output mode cleaner, squeezer beam pointing, and the arm cavities.

The interferometer must first be ``locked" to be sensitive to gravitational waves.
Locking is the process of bringing the detector into a regime where maximum power buildup is achieved in the arm cavities and all interferometer degrees of freedom are controlled \cite{Staley_2014, MartynovPhD}.
First, green lasers at each end station are locked to each arm cavity length.
Then, the green transmission beams through each arm are combined with local oscillator light on photodetectors that produce signals to control the common and differential arm cavity lengths.  Next, the power-recycling cavity, signal-recycling cavity and Michelson lengths are locked to the infrared laser via Pound-Drever-Hall error signals.

In this phase all main degrees of freedom are controlled but there is no infrared light in the arm cavities.
To transition to full infrared control, first the power-recycling, signal-recycling, and Michelson error signals are transitioned from using the first-order radio-frequency sidebands to using the third-order sidebands \cite{Arai3f}.
This is done because the first-order sideband error signals become zero as the arms are brought from antiresonance to resonance.
Then, the green common arm length is brought from infrared antiresonance to the side of an infrared fringe, where control is handed off to infrared transmission through the arms.
Next, the infrared light is brought to resonance, where both differential and common arm lengths are transferred to Pound-Drever-Hall error signals~\cite{PoundDreverHall}.
For the DC readout scheme, a \SI{10}{pm} length offset is applied to the DARM degree of freedom to allow some carrier light to leave the antisymmetric port and act as a local oscillator for light carrying the gravitational wave signal.

At this stage the entire interferometer sensing is performed via the main infrared light.
The input power is increased, low-noise controls are engaged, and squeezed light is injected to achieve maximum sensitivity to gravitational waves.
At this point the locking process is complete and the interferometer is ready for observing.

The steps taken to acquire lock are done automatically using a state machine called \textit{Guardian}~\cite{Rollins2016}.
Because the locking sequence is not deterministic and can be hindered by poor environmental conditions, there is some variability of the lock acquisition time.
The locking sequence takes approximately 30 minutes in good environmental conditions and with good initial alignment. Much of this time is used to allow various slow drift control loops to settle, allow optics to thermalize, and smoothly and reliably move between different control and actuation configurations.
Improvements to the lock acquisition are covered in Section~\ref{ss:lockAqStability}.

A ``lock loss" occurs when the detector falls out of the sensitive linear regime and control systems are unable to return to this state.
Lock losses are caused by strong earthquakes, controls and sensor saturations, drifting misalignment, control loop instabilities, and glitches of known and unknown origin.
The cause of lock losses are monitored, and if possible mitigated, to improve detector duty cycle, as described in Section \ref{ss:dutycycle}.

Section~\ref{s:overview} summarizes detector performance during O3.
Section~\ref{s:noise} describes the technical and fundamental noise sources limiting gravitational-wave sensitivity for both LIGO detectors.
Section~\ref{s:upgrades} reports the detector upgrades prior to O3.
Section~\ref{s:commissioning} discusses additional investigations at each detector.
 
\section{O3 Overview}\label{s:overview}

\subsection{Advanced LIGO noise budgets}\label{ss:noisebudgets}
Both detectors are sensitive to gravitational waves across a broad frequency band from \SI{20}{Hz} to \SI{5}{kHz}.
The sensitivity of a detector is limited by the collection of noises coupled to the gravitational-wave readout.
To improve the sensitivity of a detector, typically a source of noise or noise coupling is identified and mitigated.
The \textit{noise budget} is a tool used in this process.

\begin{figure*}[p]
  \centering
 \subfloat[LHO]{
    \includegraphics[width=0.8\textwidth]{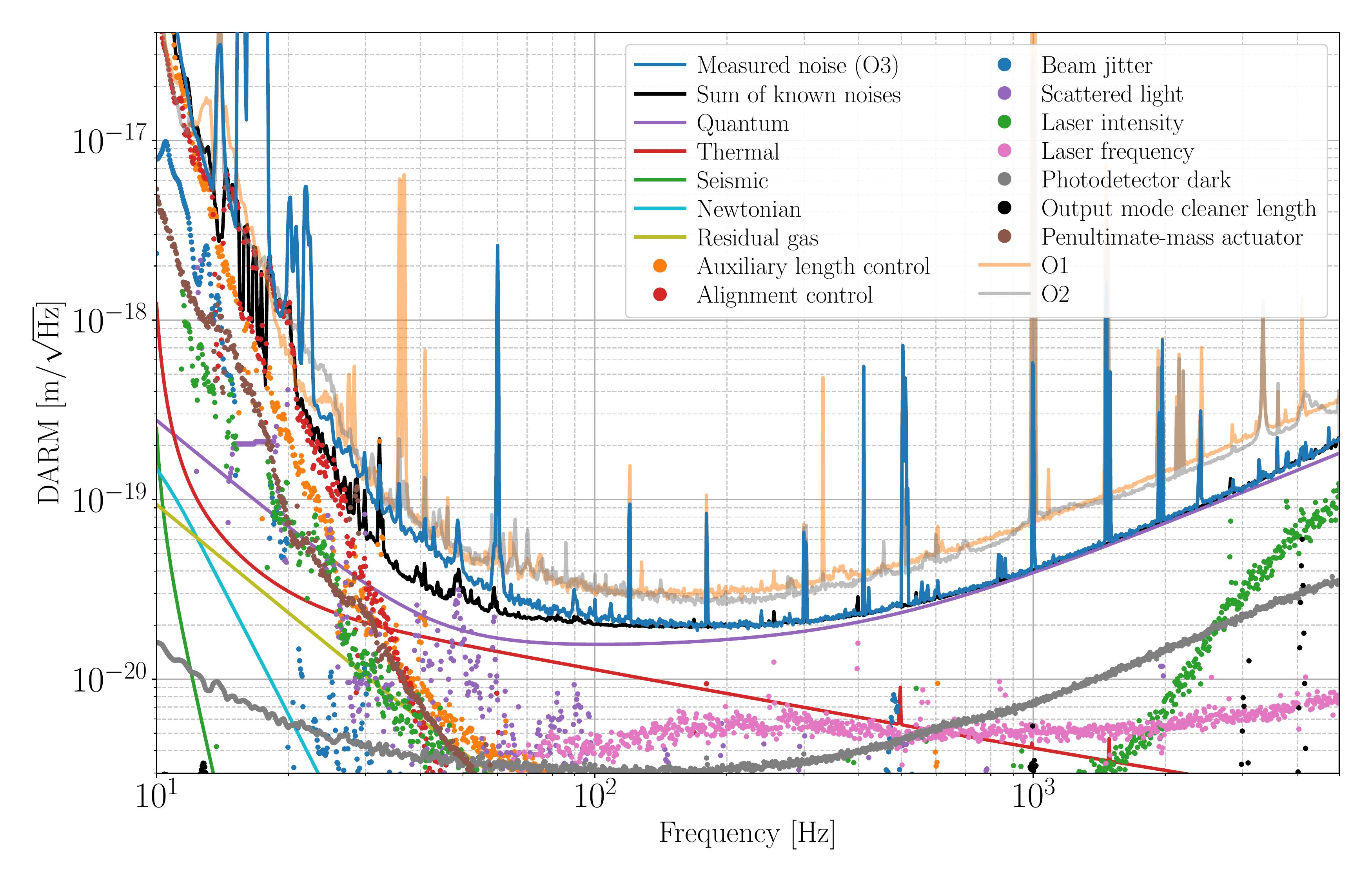}
    \label{fig:LhoNb}
    }\hfill
  \subfloat[LLO]{
    \includegraphics[width=0.8\textwidth]{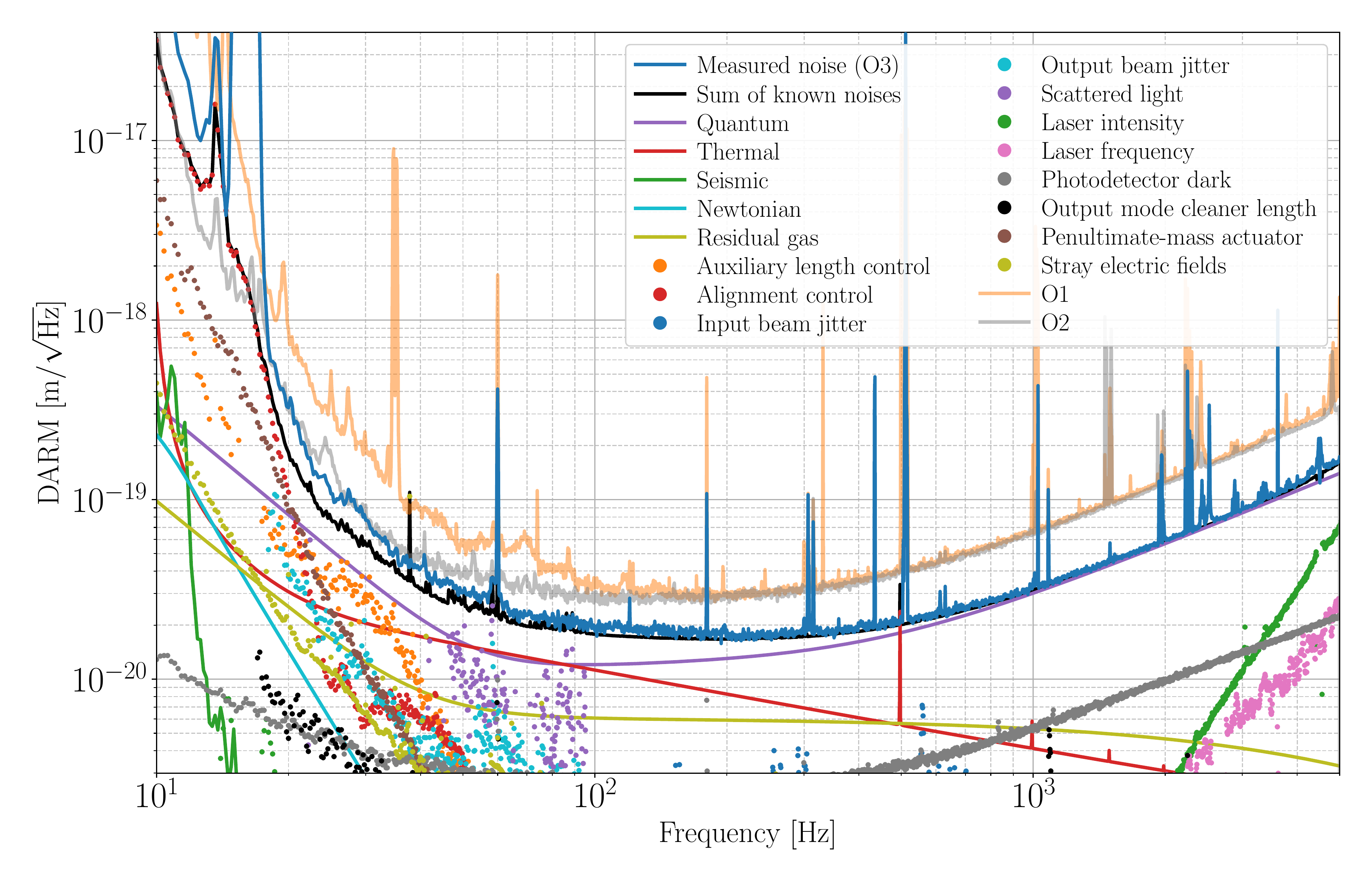}
    \label{fig:LloNb}
}
  \caption{
    Full noise budget of \protect\subref{fig:LhoNb} LIGO Hanford Observatory (LHO) and \protect\subref{fig:LloNb} LIGO Livingston Observatory (LLO).
    Calculated noise terms are given as solid lines, while measured contributions are given as dots.
    Also included are the instrument noise floors for previous observing runs, as originally presented in \cite{PhysRevD.93.112004} and \cite{Driggers:2018}.
    The O2 noise spectrum for LHO has several lines and independently witnessed noises subtracted.
    Individual noises are discussed in Section~\ref{s:noise}.
    Both sites are broadly limited by the same noise sources, with some notable differences, including beam jitter coupling (Section~\ref{ss:jitter}), laser noise couplings (Sections~\ref{ss:frequency} and \ref{ss:intensity}), and residual gas noise (Section~\ref{ss:gasnoise}).
  }
  \label{fig:nb}
\end{figure*}

The noise budget is displayed as an amplitude spectral density of the equivalent differential arm motion for the various noise sources, and is shown in Figure~\ref{fig:nb}.
Figure~\ref{fig:nb} also shows comparisons of the total measured noise in the three observing runs.
The most dramatic improvements made for O3 are due to the injection of squeezed light into the antisymmetric port (Section~\ref{ss:squeezer})
and the increase of resonating laser power inside the interferometer (Section \ref{ss:laser}), both of which improve the high-frequency sensitivity.

The noise budgets show the current understanding of the limiting noise sources at each observatory.
The black solid trace in each plot represents the sum of all known contributing noise sources.
There is excellent agreement between the modeled and measured noise above roughly \SI{100}{Hz}, but there remain additional noise sources below \SI{100}{Hz} that are not understood.
The noise budget is not used to explain narrow spectral features such as the violin resonant modes of the fused silica fibers at \SI{500}{Hz} and harmonics,
mains power at \SI{60}{Hz} and many others explained in \cite{Covas:2018}.
While the detectors are nominally physically identical, slight differences in optical properties, control loop settings, and local environments mean that the noise budgets are not identical, particularly for sources that do not limit DARM sensitivity.
We discuss each limiting noise source in Section~\ref{s:noise}.

\subsection{Astrophysical range}\label{ss:range}
Recent upgrades and improved understanding of the limiting noise sources have produced record sensitivity.
A useful metric for understanding the sensitivity of a detector is the \emph{binary neutron star inspiral range}, or simply \emph{range}.
The range reported is the luminosity distance at which a detector is sensitive to an angle-averaged merger of two \SI{1.4}{\Msol} neutron stars for a canonical SNR of 8 \cite{FinnChernoff:1993, Finn:1996, Chen:2017}.
The angle average is over the orientation of binary systems and position relative to the detector antenna patterns.
As such the range does not represent a strict maximum distance at which a binary neutron star merger can produce a significant signal.
The LIGO Livingston Observatory (LLO) has achieved a binary neutron star range of around \llorange,
while the LIGO Hanford Observatory (LHO) has achieved a range of around \lhorange.
The detector sensitivity to heavier binary black holes extends much further than binary neutron stars.

\begin{figure}
     \centering
     \subfloat[]{
         \includegraphics[width=0.95\columnwidth]{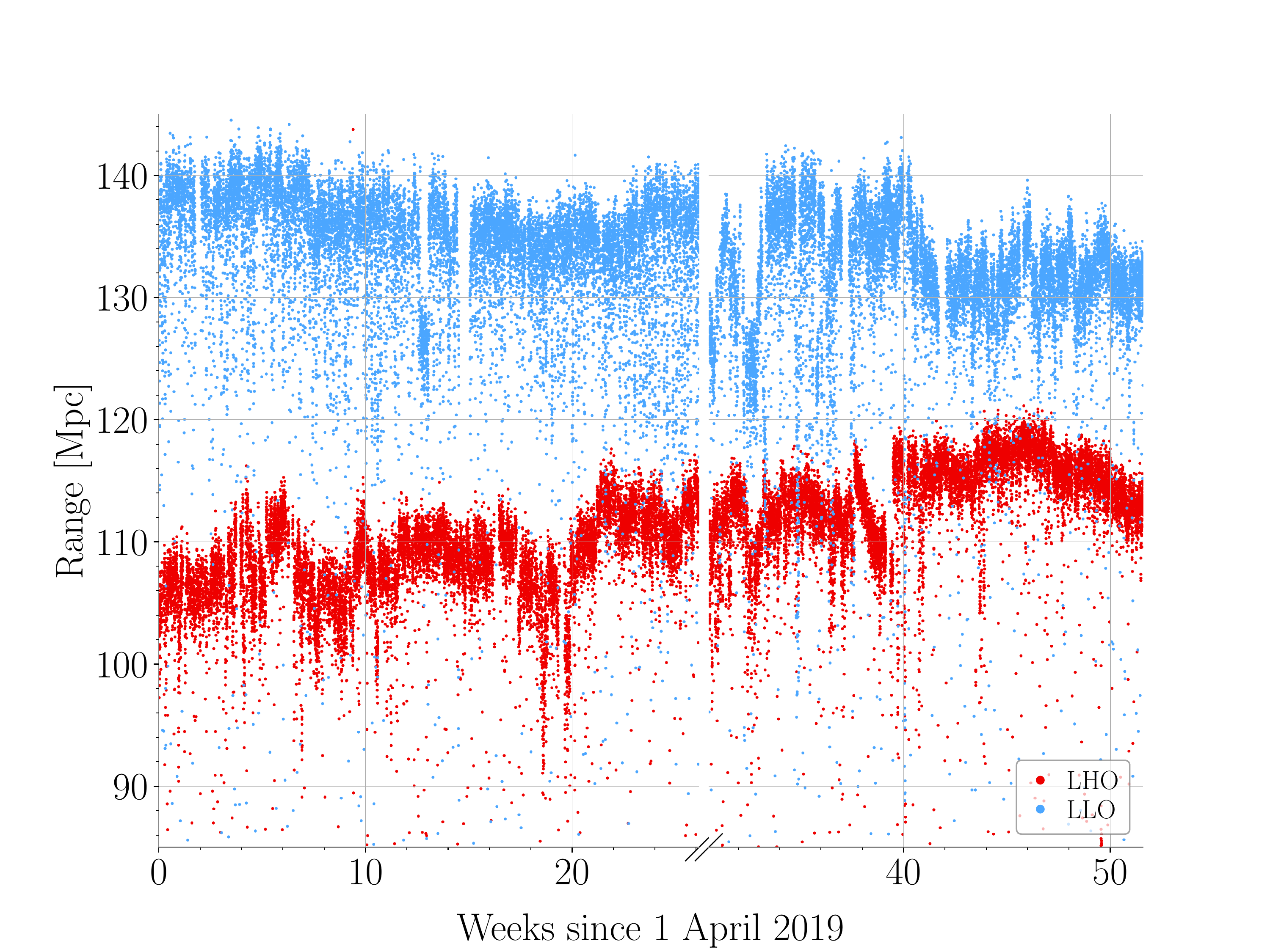}
         \label{subfig:rangevstime}
     } \hfill
     \subfloat[]{
         \includegraphics[width=0.95\columnwidth]{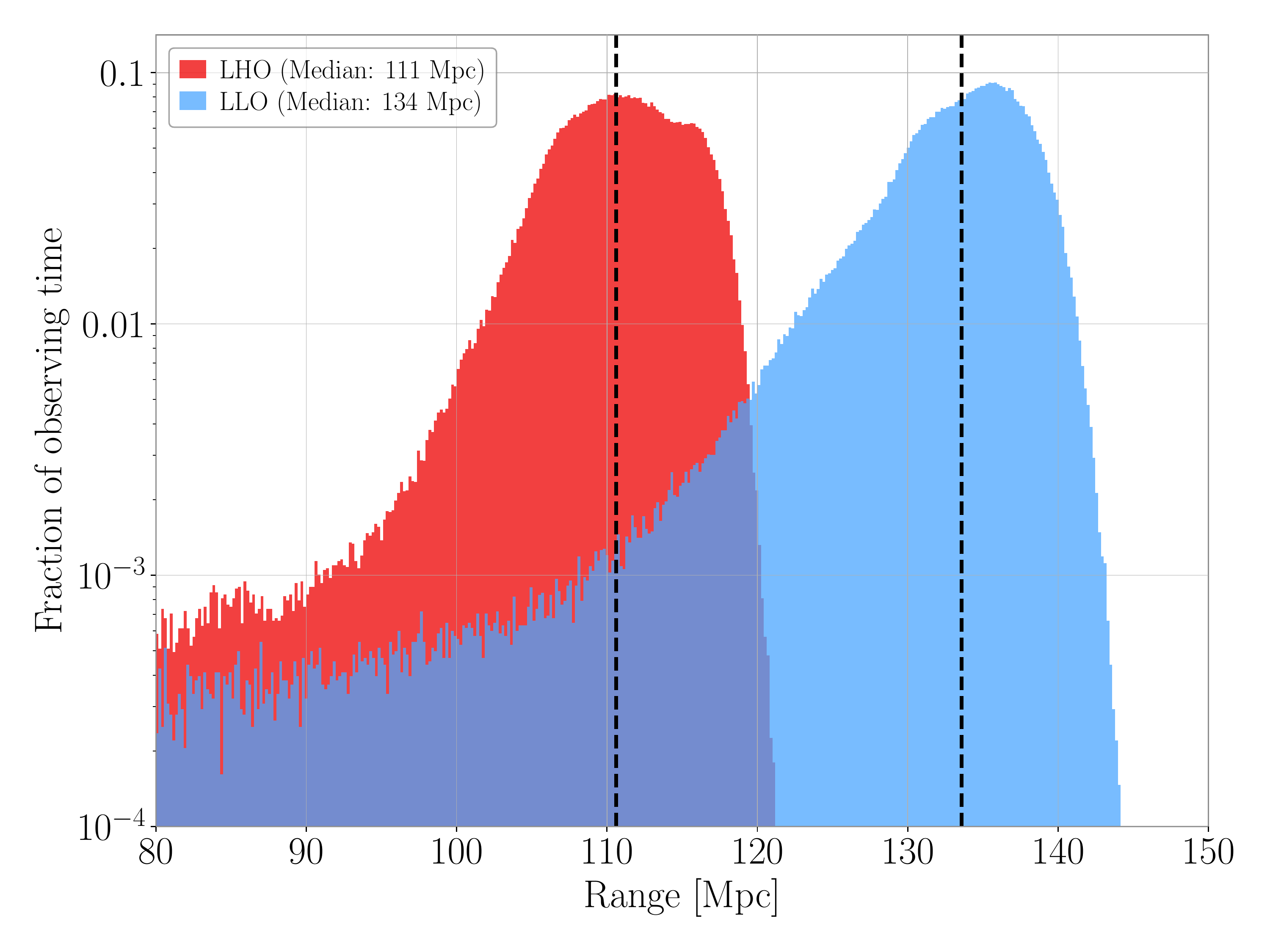}
         \label{subfig:rangehist}
     }

     \caption{The angle-averaged sensitivity of each detector (as determined by the binary neutron star inspiral range; see main text for definition) as \protect\subref{subfig:rangevstime} a function of time and \protect\subref{subfig:rangehist} a fraction of observing time.
         The time dependence is largely caused by changes in anthropogenic noise, which can increase scattered light noise.
         Additional variations are due to changes in the interferometer configuration.
         The break in the horizontal axis corresponds to the month-long observing break in October 2019.
         Brief but significant drops in the range at both sites are caused by instrumental \emph{glitches} of unknown origin (Section~\ref{ss:other}).}
         \label{fig:range}
\end{figure}

The range is calculated every minute from the online calibrated strain power spectral density;
Figure \ref{fig:range} shows the range of each observatory during O3.

\subsection{Duty cycle}\label{ss:dutycycle}
During O3 both detectors were operational a greater percentage of the time compared to the past two observing runs, with LHO and LLO achieving observation duty cycles of \lhodutySummary{} and \llodutySummary{}, respectively, with coincident observation \coincident{} of the time.
Time not observing is spent either acquiring lock, unlocked and undergoing maintenance, unlocked due to unfavorable environmental conditions (earthquakes, wind, storms), or locked and in a state of commissioning, where improvements are made to the detectors.
Improvements to the automated lock acquisition sequence, which places the detector in a detection-ready state, are outlined in Section~\ref{ss:lockAqStability}.
While some parts of lock acquisition are faster, new features have been added such that overall the lock acquisition time has not changed significantly from run to run.

Section~\ref{ss:lockAqStability} discusses improvements to detector robustness and stability that result in less frequent lock losses, longer lock durations, and improved observation duty cycle (Table~\ref{table:locking}).

\begin{table}
  \centering
  \begin{tabular}{ l | c c c c }

    Observatory & O1 & O2 & O3a & O3b \\ \hline
    \textbf{LHO} & & & \\
    \multicolumn{1}{r|}{Mean (hr)} & \num{9.8} & \num{9.4} & \num{12.4} & \num{14.9} \\
    \multicolumn{1}{r|}{Median (hr)} & \num{7.2} & \num{4.7} & \num{8.8} & \num{8.9} \\
    \multicolumn{1}{r|}{Duty cycle (\%)} & \lhoOoneduty & \lhoOtwoduty & \lhoOthreeaduty & \lhoOthreebduty \\
    \textbf{LLO} & & & \\
    \multicolumn{1}{r|}{Mean (hr)} & \num{5.7} & \num{5.5} & \num{10.2} & \num{14.5} \\
    \multicolumn{1}{r|}{Median (hr)} & \num{1.9} & \num{2.9} & \num{6.5} & \num{9.3} \\
    \multicolumn{1}{r|}{Duty cycle (\%)} & \lloOoneduty & \lloOtwoduty & \lloOthreeaduty & \lloOthreebduty \\

  \end{tabular}

  \caption{
  Mean and median times of low-noise lock segments for each observing run and overall observing run duty factor.
  Large transients or unfavorable weather and seismic conditions can knock the interferometers out of lock, reducing the total observing time.
  In addition to improved sensitivity, both detectors have improved resistance to large disturbances.
  }
  \label{table:locking}
\end{table}

\begin{figure}
     \centering
         \includegraphics[width=\columnwidth]{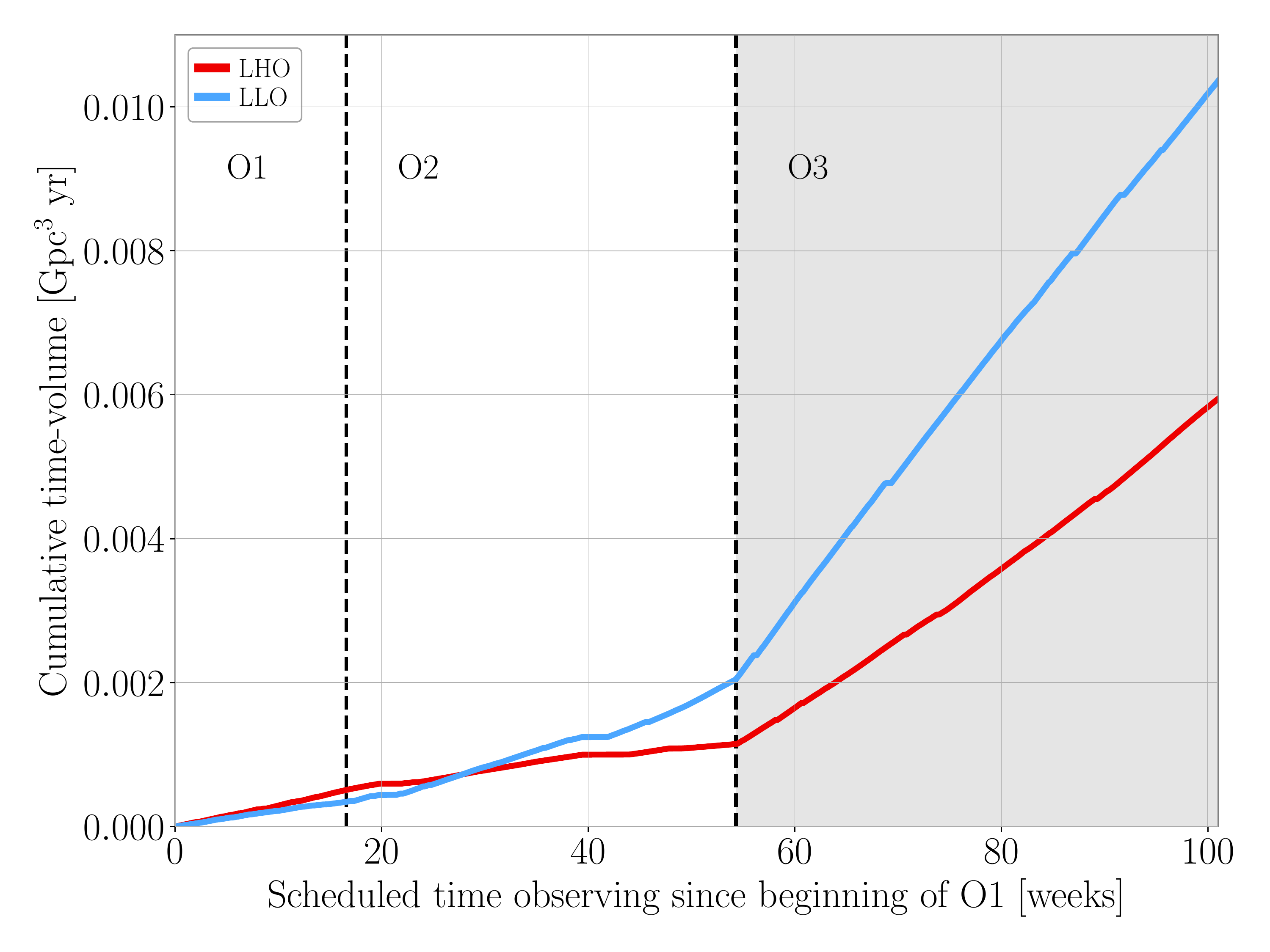}
         \caption{Integrated observation time-volume for both LHO and LLO over the durations of the three observing runs.
         The observing volume is calculated as a sphere with radius equal to the binary neutron star inspiral range, a proxy for the sensitivity of each detector (see text for definition).
         The rapid increase in this metric during O3 is due to improvements in interferometer sensitivity and duty cycle.
         The vertical dashed lines indicate the planned breaks between the three runs.
         }
         \label{fig:time_volume}
\end{figure}

Figure \ref{fig:time_volume} shows the integrated time-volume sensitivity to binary neutron stars for both sites over the three observing runs.
The increase in sensitivity combined with the higher duty factor have resulted in a dramatic increase in the observed time-volume integral, and a roughly proportional increase in gravitational-wave event candidates \cite{gracedbO3,nasaO1O2}.

\section{Analysis of instrumental noise}\label{s:noise}

Figure \ref{fig:nb} shows the current understanding of the limiting noise sources at each observatory.
These are produced using two sets of noise terms: those that are calculated based on interferometer optical and material properties~\cite{gwinc}, and projections of noises from auxiliary channels.

The most common type of projection is made by inferring a coupling function $G(f)$ between a witness channel $w$ and DARM at frequency $f$.
The contribution of noise in this channel to DARM is given by
\begin{equation}
S_{w \rightarrow d}(f) = G(f) S_{w,0}(f).
\end{equation}
Here $S_{w,0}$ is the power spectrum of witness channel $w$ under normal operating conditions, assumed not to be limited by sensing noise.
For most auxiliary channels, the coupling function is estimated according to
\begin{equation}
G(f) = \frac{S_{d,\textrm{exc}}(f)-S_{d,0}(f)}{S_{w,\textrm{exc}}(f)-S_{w,0}(f)}.
\end{equation}
Here $S_{x,\textrm{exc}}$ is the power spectrum of channel $x$ with an external excitation applied, while $S_{x,0}$ is the same spectrum in ambient conditions.
$w$ and $d$ refer to the witness channel and DARM, respectively.
$G(f)$ is set to zero where such excitations do not produce an appreciable signal in the witness channel or DARM.
Because typical witness channels have low coherence with DARM, this is \emph{not} equivalent to taking the transfer function between channels; see the example in Section~\ref{ss:LSC}.
These injections are performed at a number of different amplitudes to confirm that $G(f)$ does not depend on the amplitude of the witness signal.
However, this coupling can be modulated or up-converted from other channels and will depend on the amplitude of those signals.

Several projections presented in Figure~\ref{fig:nb} are more complicated.
In some cases, the coupling to DARM is nonlinear (such as the output mode cleaner length noise, Section~\ref{ss:omcl}).
In other cases, it can be challenging to perform an excitation that mimics and is measured by witness sensors in the same way as ambient disturbances; this is especially problematic for jitter and scattered light noise estimates (Section~\ref{ss:scatterNoise}).

What follows is a brief discussion of each noise term shown in Figure~\ref{fig:nb}.

\subsection{Quantum noise}\label{ss:quantum}
Fluctuations of the vacuum electric field at the interferometer readout port impose a fundamental limit to the interferometer sensitivity \cite{PhysRevLett.45.75, PhysRevD.23.1693, braginsky_khalili_thorne_1992, BuonannoChen2001}.
Quantum noise appears as shot noise and quantum radiation pressure noise (QRPN).

Shot noise arises from statistical fluctuations in the arrival time of photons at the interferometer output.
As the intracavity power is increased, the displacement signal-to-shot-noise ratio increases.
Shot noise dominates the high-frequency region of the spectrum.

QRPN is displacement noise arising from amplitude fluctuations of the electric field in the arms.
These amplitude fluctuations produce a fluctuating momentum on the optics via radiation pressure, inducing displacement noise.
As the intracavity power is increased, this displacement noise also increases.
QRPN is attenuated by the free-mass response of the test masses and so is more important at low frequencies.
QRPN never dominates the gravitational-wave spectrum.

In O3, shot noise is reduced by the use of squeezed vacuum injected through the antisymmetric port of the interferometer \cite{Tse2019}.
Injecting vacuum squeezed in the phase quadrature reduces the power fluctuations seen by the antisymmetric port photodiodes, lowering the shot noise floor.
However, due to the uncertainty principle, squeezing the phase quadrature leads to anti-squeezing in the amplitude quadrature, raising the QRPN floor.

The increase in laser power and installation of the squeezer has decreased the shot noise contribution.
These improvements come with a corresponding increase in QRPN, which is acceptable because QRPN does not dominate the low frequency gravitational-wave spectrum.
However, QRPN is close to limiting the current gravitational-wave noise floor~\cite{Yu2020}.

Signal-recycling mirror reflectivity also impacts quantum noise by modifying the interferometer response to gravitational waves.
The increase in signal-recycling mirror reflectivity discussed in section~\ref{ss:coreoptics} slightly broadened the region of low quantum noise while slightly increasing the minimum quantum noise.
This had a small effect on binary neutron star inspiral range.  
\subsection{Thermal noise}\label{ss:thermal}
Thermal motion in the test mass suspension, substrate, and optical coating cause displacement noise in DARM \cite{Braginsky_2003, PhysRevD.57.659, Yam2015}.
Generally thermal noise increases with mechanical loss or loss angle, as related by the fluctuation-dissipation theorem~\cite{Callen1952,Saulson1990}.
The test mass quadruple suspension system has been designed to limit thermal noise in the measurement band \cite{Aston2012}.
The fused silica substrate material is chosen for low mechanical loss and has a small contribution to the thermal noise.
A minor contribution to the thermal noise is due to the addition of acoustic mode dampers (Section~\ref{ss:parametric})~\cite{Biscans2019}.  The thermal noise contribution from these dampers is estimated to degrade interferometer sensitivity by less than 1\%.  

Brownian motion of the optic dielectric coatings is the dominant noise in the LLO noise budget from \SIrange{40}{100}{Hz}.
Advanced LIGO test masses have titania-doped tantala/silica coatings ($\mathrm{TiO}_2$-doped $\mathrm{Ta}_2\mathrm{O}_5/\mathrm{SiO}_2$), with 25\% titania in the tantala layers and varying layer thicknesses to reduce thermal noise \cite{Harry_2006,Granata_2020}.
The coating thermal noise contribution is estimated based on recent optical measurements of aLIGO end test mass witness samples \cite{GrasCoating}.
The correlated noise measurements in Section~\ref{ss:crosscor} approach the thermal noise limit as the dominant known noise source around \SI{200}{Hz}.
The coating thermal noise can be reduced with low-loss optical coatings or cryogenic optics \cite{Steinlechner2018}.

\subsection{Seismic noise}\label{ss:seismic}
The Advanced LIGO test masses form the bottom stage of a quadruple pendulum chain \cite{Aston2012}.
The purpose of this chain is to reduce coupling of ground motion (characterized at the LIGO sites in \cite{Daw_2004}) to the test mass.
These pendulums are suspended from seismic isolation platforms \cite{Matichard_2015} which themselves are supported by hydraulically actuated pre-isolation structures \cite{HEPI_2014}.

This arrangement ensures that the seismic noise contribution at the bottom of the chain sits far below the DARM noise curve.
However this seismic noise contribution only accounts for linear coupling to the DARM degree of freedom; coupling can become nonlinear when motion is large, up-converting into the gravitational-wave band.
There are circuitous paths by which seismic motion can couple to the interferometer output, such as through angular degrees of freedom (Section~\ref{ss:ASC}), auxiliary cavity length degrees of freedom (Section~\ref{ss:LSC}), or scattered light (Section~\ref{ss:scatterNoise}).
Earthquakes, high microseism, and windy conditions---which can confuse isolation systems by tilting building floors near wind-driven walls---generate additional motion that can increase scattered light coupling, cause lock loss, and hinder lock reacquisition.
 
\subsection{Newtonian noise}\label{ss:newtonian}
Newtonian noise is produced by direct gravitational coupling of test masses to fluctuating mass density fields, 
such as produced by seismicity and atmospheric pressure fluctuations \cite{Harms:2015, Driggers:2012newtonian, Coughlin:2016, Coughlin:2018}.
Newtonian noise, dominated by seismic surface waves called Rayleigh waves, is predicted to limit the design sensitivity of the Advanced LIGO detectors from \SIrange{10}{20}{Hz} \cite{Saulson:1984, Hughes:1998}.
Newtonian noise has not been detected in Advanced LIGO, and is predicted to be below O3 sensitivity levels \cite{PhysRevD.101.102002}.
 
\subsection{Laser frequency noise}\label{ss:frequency}
Frequency noise refers to the fluctuations in the instantaneous frequency of the laser.
Frequency noise can appear as differential phase fluctuations in the arms, masking the gravitational-wave signal.

The common-mode rejection of the interferometer ensures most frequency noise does not reach the antisymmetric port.
Asymmetries in the interferometer allow frequency noise to appear at the dark port,
including the DARM offset and Schnupp asymmetry purposefully introduced for interferometer control,
and unintentional mismatch in the arm reflectivity, losses, cavity pole, power buildup, and transverse mode content \cite{Izumi_FreqRespPart3_2015, Somiya:2006kb, Camp:2000ht, Cahillane:2020}.

A frequency stabilization servo (FSS) is employed both to lock the laser to the extremely narrow common-arm linewidth and to suppress the free-running frequency noise of the main laser \cite{Fritschel:01}.
There are three hierarchical control loops.
The first stage is the reference cavity, a \SI{20}{cm} fixed-length cavity to which the laser frequency is initially stabilized \cite{Kwee2012}.
The second stage is the input mode cleaner, where the laser is stabilized to the \SI{33}{m} suspended cavity.
The third stage is the common-arm cavity, where the laser is stabilized to the \SI{4}{km} arm length, with coupled cavity pole of \SI{0.6}{Hz}.
All three stages together suppress the frequency noise so it does not limit the gravitational-wave spectrum.

Two upgrades to the frequency stabilization loop were performed at LIGO Hanford.
First, a second photodetector was added to detect the light reflected from the interferometer, where the common-arm length degree of freedom is sensed.
This allows for the two photodetectors to be used in an in-loop, out-of-loop configuration to directly measure the sensing noise in the loop.
During operation, both photodetectors are used in-loop to reduce the risk of saturations.

\begin{figure}
  \centering
  \includegraphics[width=\columnwidth]{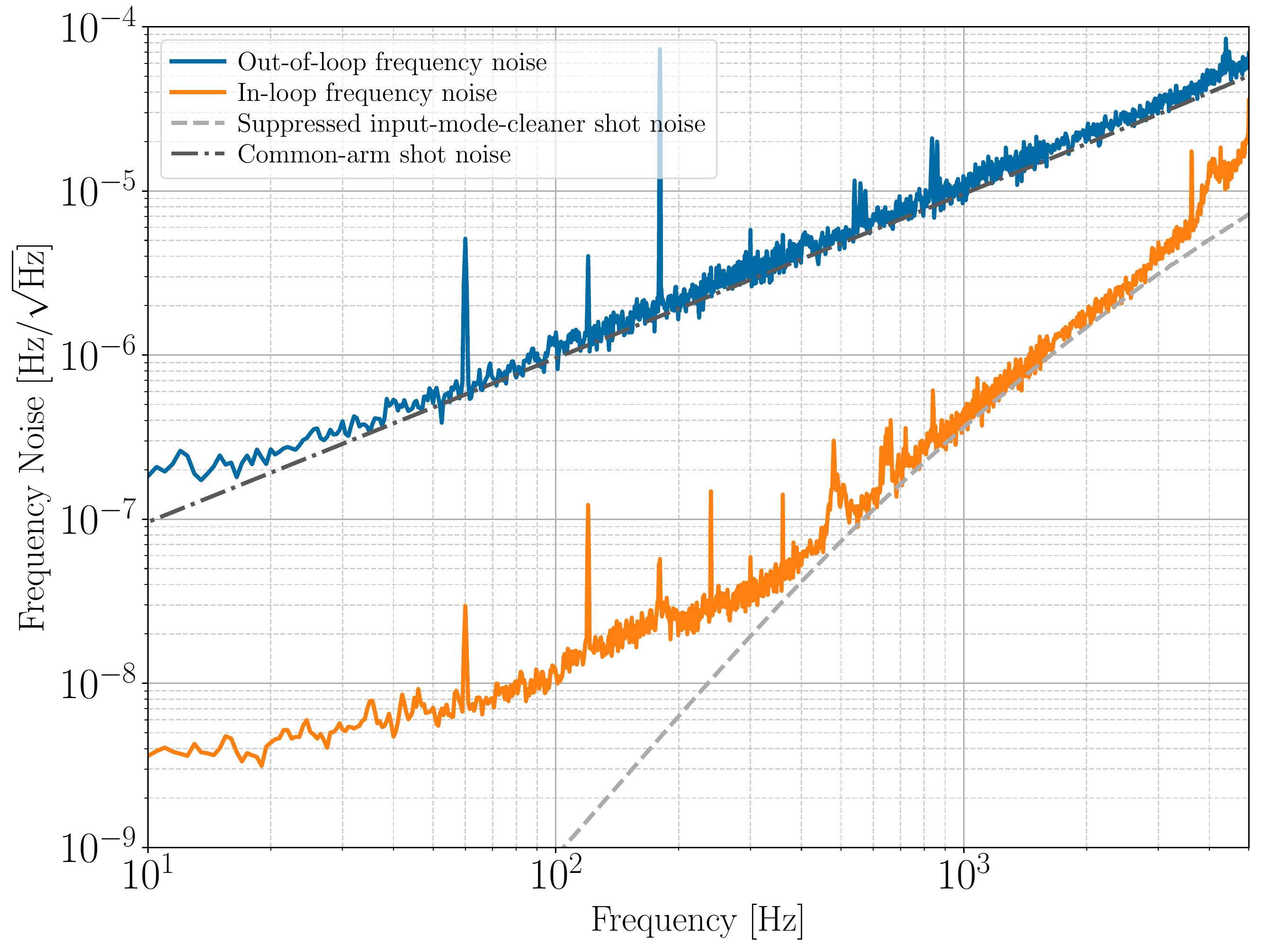}
  \caption{
  LIGO Hanford laser frequency noise budget. The upper trace is the out-of-loop witness of frequency noise incident on the interferometer, which is dominated by shot noise.
  This trace is projected into the gravitational-wave spectrum in Figure~\ref{fig:LhoNb}.
  The in-loop frequency noise is limited by shot noise from the input mode cleaner (lower trace).
  During operation, the reflection shot noise is a factor of 2 lower than plotted here
  as both reflection photodiodes are in-loop, doubling the common-arm length signal.
  }
  \label{fig:frequencyNoisebudget}
\end{figure}

Second, the power on the input mode cleaner reflection photodetector was increased by a factor of seven.
This improved the optical gain of the second stage of the FSS, reducing the frequency noise incident on the interferometer.
Figure \ref{fig:frequencyNoisebudget} shows the current frequency noise budget after these changes.

Frequency noise couplings to the gravitational-wave spectrum were found to change significantly with the thermal state of the interferometer, likely due to changing transverse mode content \cite{Cahillane:2020}.
These couplings were partially mitigated via the thermal compensation system (Section \ref{ss:tcs}).

\subsection{Laser intensity noise}\label{ss:intensity}
Intensity fluctuations of the laser appearing at the interferometer dark port can mask gravitational-wave signals.
Similar to frequency noise, the common-mode rejection of the interferometer is not enough by itself to avoid sensing intensity fluctuations in DARM.
Advanced LIGO employs an intensity stabilization servo (ISS) made of two hierarchical control loops to suppress the laser intensity fluctuations incident on the interferometer.
Both ISS loops feed back to a single-pass acousto-optic modulator (AOM) that actuates on the laser power.

The ISS first loop stabilizes the total laser power out of one port of the bow-tie pre-mode cleaner on the pre-stabilized laser table with a bandwidth of \SI{80}{kHz}.
The second ISS loop stabilizes the total power transmitted through the input mode cleaner cavity.
A pick-off of this cavity transmission goes to an in-vacuum array of eight photodiodes.
Four of the eight photodiodes are used for the second loop sensor.
This control signal is filtered and summed with the control signal from the first loop and sent to the AOM.
The other four photodiodes are out-of-loop witnesses of intensity noise.
The ISS second loop has a bandwidth of around \SI{28}{kHz}.

\begin{figure}
  \centering
  \includegraphics[width=\columnwidth]{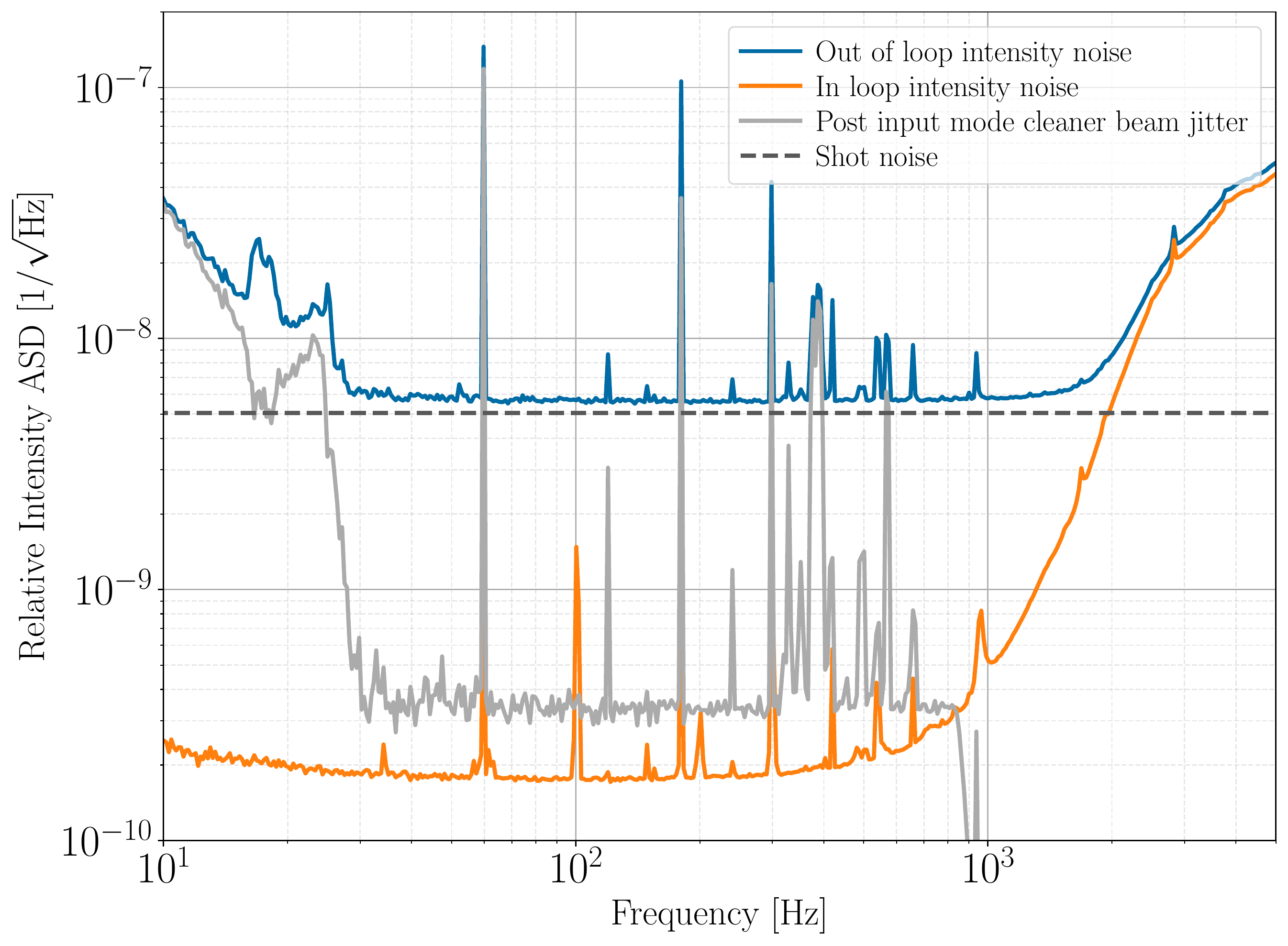}
  \caption{
  LIGO Hanford laser intensity noise budget.
  The uppermost trace represents the out-of-loop witness of intensity noise incident on the interferometer.
  This trace is projected into the gravitational-wave spectrum in Figure~\ref{fig:LhoNb}.
  Seismic motion causes intensity fluctuations below \SI{1}{Hz}, and input mode cleaner suspension resonances dominate the intensity RMS.
  Beam jitter dominates from 4 to \SI{30}{Hz},
  the shot noise limit is attained from 30 to \SI{1000}{Hz},
  and the intensity stabilization servo is gain-limited above \SI{1}{kHz}.
  }
  \label{fig:intensityNoisebudget}
\end{figure}

Figure~\ref{fig:intensityNoisebudget} illustrates the laser intensity stability at Hanford.
Intensity noise RMS is dominated by input mode cleaner angular control peaks between 1 and \SI{4}{Hz}.
Between 4 and \SI{30}{Hz}, jitter in the beam path after the input mode cleaner causes apparent intensity fluctuations on the ISS photodiode array, as witnessed by the ISS quadrant photodiode.
The shot noise floor of the second loop is attained between \SI{30}{Hz} and \SI{1}{kHz}.
Unsuppressed intensity noise dominates above \SI{1}{kHz}, where the intensity servo is gain-limited.

The large difference between the Hanford and Livingston intensity noise contributions to DARM at low frequencies, as seen in Figure~\ref{fig:nb}, 
is likely due to an increased radiation pressure coupling due to a larger arm power mismatch at Hanford.
The circulating arm powers for both sites are discussed in Section~\ref{ss:ArmPowerMeas}. 

\subsection{Auxiliary length control noise}\label{ss:LSC}
The gravitational-wave readout is orders of magnitude more sensitive to differential arm cavity length (DARM) than to the lengths of the power-recycling cavity, signal-recycling cavity, or the Michelson degree of freedom (beam splitter position relative to the arm input mirrors).
However, each of these auxiliary degrees of freedom must be controlled to keep the interferometer in its sensitive configuration, and DARM readout is still marginally sensitive to each.
Each has an individual readout scheme \cite{Izumi2017} that is less sensitive than the main DARM readout.

The gains of each auxiliary loop are chosen to be high enough to always control the interferometer while being as low as possible to minimize re-injected sensing noise.
Even with this gain optimization, sensing noise from the Michelson and signal-recycling cavity loops can couple to DARM and would limit interferometer sensitivity if not for feedforward cancellation.

\begin{figure}
  \centering
  \includegraphics[width=\columnwidth]{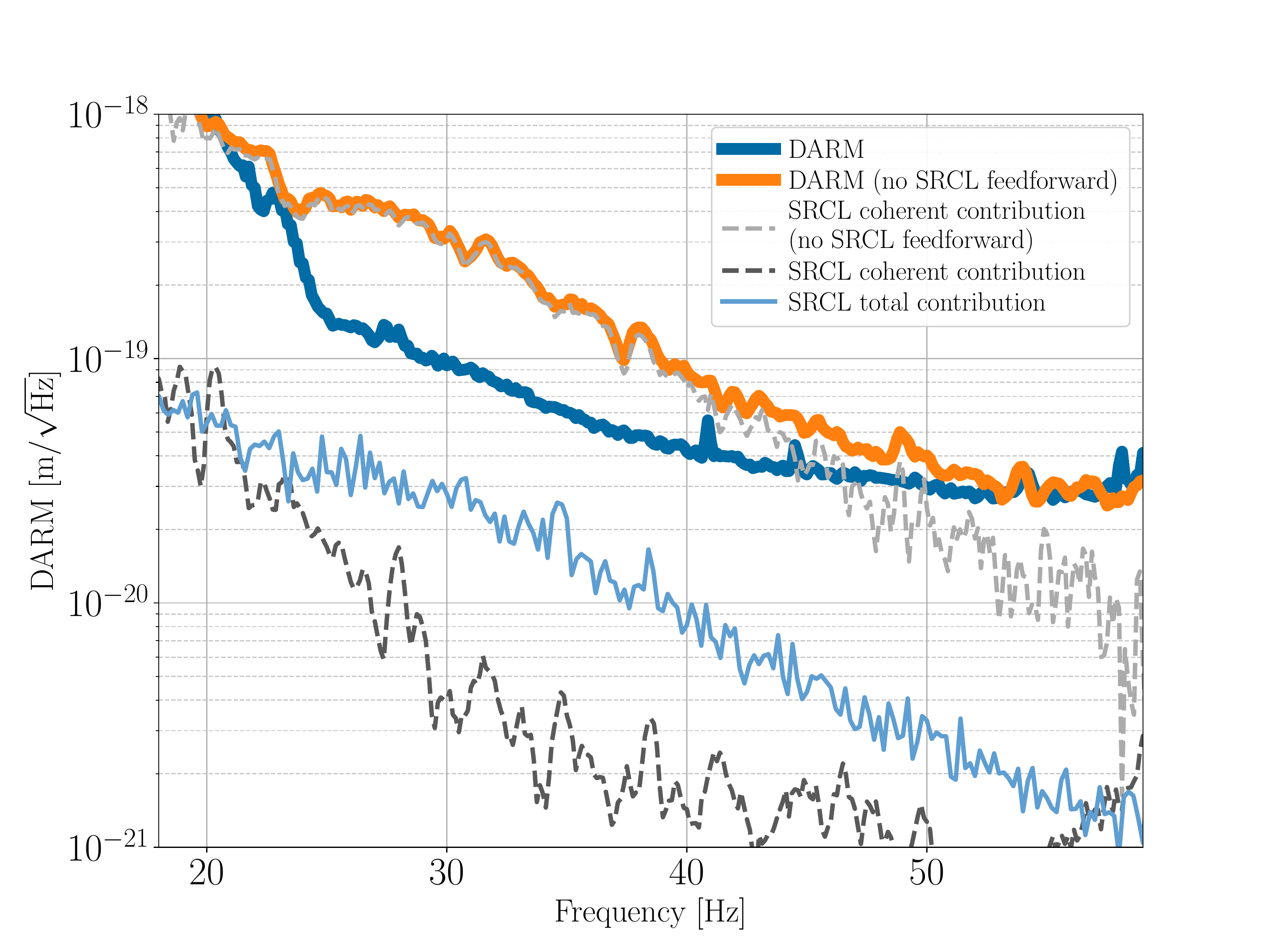}
  \caption{
  The contribution of the signal-recycling cavity length (SRCL) control signal to the gravitational-wave readout channel (DARM) at LLO.
  The coherent contribution is determined by the magnitude-squared coherence between DARM and SRCL, while the total contribution is estimated as described at the beginning of Section~\ref{s:noise}.
  By using a feedforward signal from the SRCL control to DARM, the coherent contribution is significantly reduced, but there remains a component that cannot be removed with simple feedforward.
  }
  \label{fig:srcl}
\end{figure}

Feedforward is a technique of real-time noise subtraction in DARM.
Re-injected sensing noise in the auxiliary length loops is measured and known to limit DARM.
The transfer functions from the auxiliary loop to DARM and the feedforward actuation path to DARM are first measured and fit using software tools such as IIRrational~\cite{iirrational}.
The output signal of this auxiliary loop is injected into DARM with this transfer function and opposite sign, cancelling the auxiliary noise that normally would appear in DARM.
Such feedforward loops have reduced the magnitude-squared coherence between these channels and DARM to below $10^{-2}$ above \SI{10}{Hz}.
Figure \ref{fig:srcl} shows how a feedforward filter between the signal-recycling length control signal and DARM reduces the coherent contribution of this noise source to DARM, defined as \cite{bendat2011random}
\begin{equation}
\frac{\left|S_{wd}(f)\right|^2}{S_{w}(f)} = \gamma_{wd}^2(f) S_{d}(f),
\end{equation}
where $S_{w}(f)$ and $S_{d}(f)$ are the power spectral densities of the auxiliary (witness) channel and DARM, respectively, and $\gamma_{wd}^2(f)$ and $S_{wd}(f)$ are the magnitude-squared coherence and cross-spectral density between these channels, respectively.

As seen in Figure~\ref{fig:srcl}, the contribution of these auxiliary channels to the DARM noise is larger than expected based on coherence alone, suggesting nonlinear, bilinear, and/or non-stationary coupling to DARM.
Non-stationary coupling has already been observed due to modulation from motion of the angular degrees of freedom, and can be partially removed offline~\cite{vajente2019machinelearning,geo2020bilinear}.
Additional work is required to understand this type of contribution to the interferometer noise floor.
 
\subsection{Actuator noise}
The position of the LIGO optics is controlled with digital-to-analog converters (DACs) and either electromagnetic coils or electrostatic actuators.
Analog electronics filter the output of the DACs and allow conversion between a high-range configuration for lock acquisition and a low-noise configuration for normal operation.
Upper suspension stages have larger actuation range but due to the suspension response have less control authority---and therefore, lower noise coupling---at frequencies above the pendulum response \cite{Aston2012}.
As such, the test mass and penultimate mass actuators are most important for direct noise coupling in the gravitational-wave band.

Operating with higher actuator range minimizes instrument susceptibility to saturations and lock loss, which can significantly negatively affect observing duty cycle.
However, this generally comes at the cost of increased noise injection.
Improvements to the actuators and digital-to-analog converters have helped both to move this noise contribution safely below the current sensitivity and improve duty cycle.
These are discussed in Section~\ref{ss:dac}.
 
\subsection{Alignment control noise}\label{ss:ASC}
The alignment sensing and control (ASC) system controls the alignment of interferometer optics.
The mirrors must be actively aligned to suppress motion from external disturbances, maximize optical power coupling,
and counteract instabilities from radiation pressure \cite{Dooley_2013}.
During lock acquisition, large increases in optical power result in radiation pressure that can push the optics out of alignment.
During low-noise operation, slow drifts in alignment must be corrected and radiation pressure torques on the optics must be compensated to maintain stable operation \cite{advancedLIGO}.

Below \SI{25}{Hz} the angular arm controls are the largest known source of noise contribution to DARM.
As described in \cite{Barsotti2010} and in \cite{PhysRevD.93.112004},
any residual angular motion is expected to couple to the longitudinal degrees of freedom of the cavities through both static and dynamic mis-centering of the beams,
leading to linear and nonlinear coupling.

Centering the beams on the suspensions by adjusting the spot position on the optic or by digitally compensating
the position of the rotation point is a critical step in reducing the ASC noise contribution to DARM.
The centering of the beams is discussed in Section~\ref{ss:alignmentdithersystem}.

Changes in cavity alignment are primarily sensed with dedicated interferometric sensors called wavefront sensors.
These quadrant photodetectors, with radio-frequency demodulation on each segment, rely on the relative alignment of carrier and sidebands inside the interferometer \cite{Anderson1984, Morrison:94}.
The sensing of the residual angular motion above \SI{10}{Hz} is limited by the noise of these sensors.
To filter this noise and achieve low-noise operation, aggressive low-pass filters in the ASC control loops are engaged.
This critical step in the lock acquisition sequence reduces the angular control gain above \SI{10}{Hz},
and therefore produces orders of magnitude reduction in angular control noise coupling to DARM.
However, this also reduces the phase margin of the loops to close to few tens of degrees.
While acquiring lock, the low-pass filters are not engaged: the ASC loops are operated with larger phase margins to cope with large radiation pressure transients.
More details on the ASC control scheme are given in Section~\ref{ss:radiationpressurecompentation}.

The wavefront sensors can also be affected by spurious local noise coupling.
At both sites, the sensing of one of the arm common angular modes is contaminated by the vertical motion of the in-vacuum table where the sensors are located.
A feedforward scheme has been implemented at LLO that reduces the impact of this effect.
At LHO, the wavefront sensors signals are blended with local reference sensors (quadrant photodiodes) in the transmission of the arms that are free of this coupling.

In the low-noise configuration, the contributions from ASC are the dominant known source of noise below \SI{25}{Hz}.
The contribution is currently smaller at LLO than LHO.
The coherence between the ASC signals and DARM is low, suggesting that this coupling is primarily nonlinear.
Upgrades to the ASC system for O3 are discussed in Section~\ref{ss:angular} and high-power alignment control issues are discussed in Section~\ref{ss:radiationpressuretorque}.
 
\subsection{Beam jitter noise}\label{ss:jitter}
Alignment fluctuations of the beam at the interferometer input couple additional noise to DARM via the changing coupling of the fundamental optical transverse mode to the arm cavities \cite{Mueller2005, HardwickPhD}.
The beam-position-dependent absorption introduced by point absorbers on the input test masses (Section~\ref{ss:absorption}) is also believed to couple jitter noise to DARM by breaking the symmetry of the arms.

During the first two observing runs, the most severe jitter noise originated from the LHO pre-stabilized laser, where vibration from the water cooling system of the high-power oscillator produced multiple peaks between 100 and \SI{900}{Hz} in DARM.
These peaks were associated with resonances of optic mounts on the pre-stabilized laser table, which are identified by individually exciting each mount while monitoring spectra of beam jitter sensors.

Before O3 the high-power laser oscillator was replaced (see Section~\ref{ss:laser}), allowing a reduction in cooling water flow.
In addition, several optical mounts on the pre-stabilized laser table were mechanically damped.
Removal of the high-power oscillator is also thought to be responsible for reduction of the broad peak between 250 and \SI{800}{Hz} by reducing fluctuations in beam size.

\begin{figure}
  \centering
    \includegraphics[width=\columnwidth]{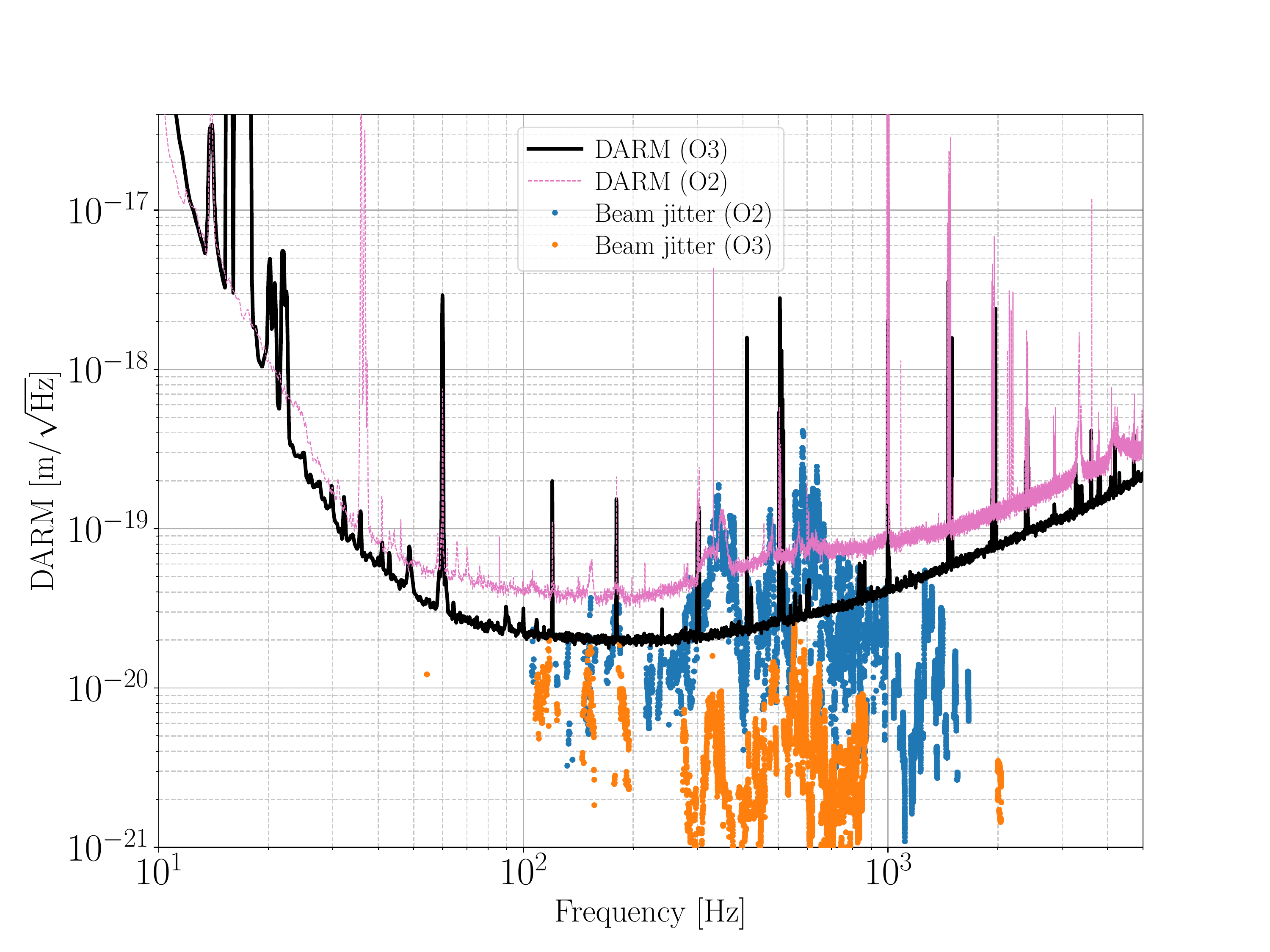}
  \caption{Jitter noise from the pre-stabilized laser, measured at LHO.
  Data from O2 (O3) is the upper (lower) dotted trace.
  Current total displacement sensitivity (solid) is compared to the O2 displacement sensitivity (dashed).
  Peaks in the spectrum are mechanical resonances coupled to DARM via beam jitter.
  At some frequencies jitter noise is over- or under-estimated because the accelerometers used to compare ambient and injected motion are near but not on the optics that produce the beam jitter.}
  \label{fig:Jitter}
\end{figure}

Figure~\ref{fig:Jitter} shows the jitter noise improvement between O2 and O3 in the LHO DARM spectrum as well as in estimates of the angular jitter noise contribution to DARM based on acoustic and mechanical vibration injections at the pre-stabilized laser table.
The most severe peaks around \SI{300}{Hz} at LHO show a reduction in amplitude of approximately an order of magnitude.
The jitter noise was lower at LLO during O2, where the high-power oscillator was bypassed \cite{HardwickPhD}.

There are several remaining jitter peaks at LHO and LLO which are just below DARM and originate from within the pre-stabilized-laser room.
Further reductions in water flow may reduce their amplitude.
Note that the linearly coupled component of jitter can be removed in post-processing, as in the LHO O2 trace in Figure~\ref{fig:nb} \cite{Driggers:2018}.

Beam jitter also couples to DARM at the output port of the interferometer.
Fluctuations in beam pointing change the coupling of the beam to the output mode cleaner (OMC)~\cite{omc_design}.
At LLO the beam pointing to the output mode cleaner is controlled with an angular dither alignment scheme.
In this scheme, modulation is applied to the actuators that control the four degrees of freedom of beam pointing to the OMC.
The modulation frequencies used in O3 were between \SI{2.2}{kHz} and \SI{2.3}{kHz}.
Light transmitted through the OMC is then demodulated to produce the alignment error signals.
The control loop bandwidths are less than \SI{10}{Hz}.
Residual motion from this control is projected to DARM with a technique that uses the root mean square level of beam motion to provide the linear coupling factor.
Output jitter projections at LLO are estimated using the quadrature sum of the contributions from the four angular degrees of freedom.

At LHO the beam pointing to the OMC is controlled with a quadrant photodetector scheme.
The beam is actively aligned onto two quadrant photodetectors on a pick-off of the light incident on the OMC.
The position of the beam on the quadrant photodetectors is chosen to maximize the interferometer optical gain.
However, for one of the alignment degrees of freedom a different beam position gave the lowest coupling of OMC angular control noise to DARM.
This suggests some unwanted light is incident on the OMC.
The projection from the OMC angular control noise to DARM at LHO was not made for O3. 
\subsection{Scattered light noise}\label{ss:scatterNoise}
When light lost from the main interferometer beam reflects or scatters off a moving surface it acquires a time-dependent phase shift.
If this scattered light re-couples to the main interferometer beam it will introduce noise to DARM \cite{Billing_1979}.

When the displacement of the scattering surface is small relative to the wavelength of light, the scattered light noise couples linearly \cite{Vinet96}.
When the displacement is much larger than the wavelength, scattered light coupling is highly nonlinear.
In this large-motion regime, known as fringe wrapping, the noise is approximately flat in frequency, with an amplitude proportional to the intensity of the scattered beam and maximum frequency related to the speed of the scattering object as described in Section~8 of \cite{MartynovPhD}, and in \cite{Ottaway12}.

During times of high ground motion, fringe wrapping can significantly degrade detector sensitivity.
This is apparent in Figure~\ref{fig:range} where LLO was more susceptible to scattered light for the first 30 weeks, resulting in greater variability in the binary neutron star inspiral range.
Mitigation efforts subsequently reduced this variability~\cite{Soni2020}.
As discussed further in Section~\ref{ss:scatterUpgrade}, baffles, mechanical damping, reaction chain actuation, and transmission monitor suspension actuation were implemented before and during O3 to mitigate stray light noise.
The contribution in Figure~\ref{fig:nb} represents the scattered light noise in times of typical ground motion and does not include the contribution from ground motion up-conversion during times of high microseism.

\subsection{Residual gas noise}\label{ss:gasnoise}
Residual gas in the vacuum chambers adds noise in two ways: as additional phase noise due to the traversal of gas molecules across the arm cavity laser beam path \cite{Zucker1996}, and as damping of the test masses \cite{PhysRevD.84.063007}.
The latter contribution has been estimated to be significantly reduced following the installation of the annular end reaction masses (Section \ref{ss:coreoptics}).

A small intermittent vacuum leak appeared near the mid point of the X arm at LLO several years ago, which increased the pressure from a few nanotorr to a few tens of nanotorr.
The correlated noise measurement at LLO in Section~\ref{ss:crosscor} is consistent with a contribution from excess gas noise.
The contribution of this noise depends on the partial gas pressure at different points along the arms.
Large uncertainties in vacuum gauge readouts, poor spatial resolution, and uncertainty of the residual gas constituents make it challenging to estimate the induced phase noise along the length of the beam.

Two leaks were located in October 2019 using standard helium leak checking techniques.
The leaks appeared at corroded areas that show visual similarity to microbial-induced corrosion \cite{MIC_Little}.
Such corrosion may have occurred in the humid environment formed by rodents and insulation that surrounded the pipe that was installed for the initial bake to remove residual water in the tube inner surface \cite{david2019advanced}.
This insulation was removed in 2017.

After leak repair, residual gas pressure at the mid station of the X-arm returned to a few nanotorr.
For the latter half of O3, the LLO residual gas contribution is expected to be reduced to the LHO contribution.
 
\subsection{Photodetector dark noise} \label{ss:dark}
Photodetector dark noise refers to the noise on the gravitational-wave photodetection chain when there is no light on the two diodes on transmission of the output mode cleaner.
This incorporates the dark noise of the diodes as well as the associated electronics.
Dark noise is a technical noise source, roughly a factor of 5 below DARM.
 
\subsection{Output mode cleaner length noise} \label{ss:omcl}
The output mode cleaner (OMC) is a bow-tie cavity that transmits the fundamental interferometer mode that carries the DARM information while reflecting higher-order transverse modes and modulation sidebands.
This reflected light has relatively large phase noise and intensity noise, as it contains light not filtered by the arm cavities.
Length fluctuations of the OMC cause fluctuations of the transmitted power that introduce noise to the gravitational-wave readout.

The OMC length is controlled with a dither lock scheme.  OMC length modulation is applied at \SI{4.5}{kHz} at LLO and \SI{4.1}{kHz} at LHO with a piezoelectric actuator on one OMC cavity mirror.  The signal from the transmitted light is demodulated to produce an error signal used to control the cavity length via the same actuator.  The control scheme is designed to have an OMC length noise of \SI{3e-16}{m/\rtHz}, safely below DARM~\cite{omc_design}.

The OMC length noise is more than a factor of 10 below DARM at both detectors.
There are small contributions around the frequency of the dither line, at injected calibration lines, and mechanical resonance lines.
There is also a small low-frequency noise contribution.
Differences in this contribution at LLO and LHO are likely due to differences in the OMC length control schemes.   
\subsection{Other}\label{ss:other}
The understanding of other noise sources has not changed dramatically since previous observing runs, though the difference between measured DARM noise and the sum of known noises has significantly decreased since O1.
Narrow spectral features are mostly understood and are either the electrical mains (\SI{60}{Hz} and harmonics), single-frequency excitations for control signals or calibration (e.g., 10--\SI{20}{Hz}, \SI{410/435}{Hz}, and \SI{1083}{Hz}),
or suspension violin-mode resonances (\SI{\sim 300}{Hz}
and harmonics for beam splitter, \SI{\sim 500}{Hz} and harmonics for test masses).
These narrow spectral features do not appreciably affect compact binary coalescence detection, although they can affect searches for continuous gravitational-wave sources~\cite{Covas:2018}, and at sufficiently high amplitude can introduce nonlinear effects.

Large transients in the gravitational-wave channel are still observed regularly, affecting the sensitivity as seen in Figure~\ref{fig:range}.
Such transients---also called ``glitches" \cite{Powell2017,Zevin2017,Cabero2019}---reduce the amount of clean data, decrease the significance of real gravitational wave signals, and, if they occur during a real signal, can complicate parameter estimation \cite{PhysRevD.98.084016}.
While there has been progress in reducing whistle glitches (Section~\ref{ss:vco}), the causes of other types of glitches are poorly understood.
Preliminary evidence suggests that \emph{tomte}-type glitches may be caused by charge transfer on the high-voltage ESD actuators (Section~\ref{ss:charging}).
  
\section{Instrument improvements}
\label{s:upgrades}

This section will discuss the instrument upgrades that facilitated the increase in sensitivity and duty cycle for O3,
focusing on hardware upgrades to the interferometers.

\subsection{Laser power increase}
\label{ss:power}
Increasing the laser power reduces instrument noise at high frequency
where the sensitivity is shot-noise-limited but comes with operational challenges.
Hardware upgrades to the pre-stabilized laser and core optics
allowed for an increase in average circulating power in the arm cavities to \lhoAveArmPower{} at LHO and \lloAveArmPower{} at LLO for O3 (see Table~\ref{table:ArmPowers}).

The major technical challenges of operating a high-power interferometer are caused by radiation pressure inducing instabilities and absorption of the test masses.
These instabilities are discussed in Section~\ref{ss:parametric} and the angular controls system is discussed in Section~\ref{ss:angular}.
Thermal distortion of the test masses due to optical absorption is discussed in Section~\ref{ss:tcs}.

    \subsubsection{Laser hardware changes}
    \label{ss:laser}
    The original aLIGO pre-stabilized laser (PSL) design \cite{Kwee2012} took the output of a Nd:YAG non-planar ring oscillator (NPRO) operating at \SI{1064}{nm} and successively amplified the output to above \SI{150}{W}.
The original amplifier chain consisted of a \SI{35}{W} solid-state amplifier (``front end") followed by a high-power injection-locked ring oscillator.
In addition to operational challenges, the latter high-power oscillator and its high coolant flow produced fluctuations of the beam size and pointing angle \cite{HardwickPhD}.
This beam jitter noise is further discussed in Section~\ref{ss:jitter}.

For O3 the high-power oscillator was replaced at both observatories with a smaller single-pass solid-state amplifier (neoLASE neoVAN-4S) that requires less coolant flow.
The new amplifier produces roughly \SI{70}{W} of stable output power during the run.
After input optics and mode-cleaning cavities, this provides up to \SI{50}{W} at the power-recycling mirror.

The reduced coolant flow and damping and tuning of problematic optic mounts has reduced the amplitude of angular beam jitter.
The higher input power, in addition to the squeezer (Section~\ref{ss:squeezer}), lead to improved sensitivity above \SI{100}{Hz}.

    \subsubsection{Parametric instabilities}
    \label{ss:parametric}
    High circulating power in the arm cavities can excite the internal acoustic modes of the test masses via radiation pressure.
When a test mass acoustic mode overlaps with a higher-order optical mode, light can be scattered into this higher-order mode.
This will further amplify the mechanical motion, increasing the scatter rate, eventually becoming a runaway process.
This is known as a parametric instability, and has been previously observed at both sites \cite{Evans2010,Evans2015}.

Before O3, acoustic mode dampers were installed on all test masses to mitigate parametric instabilities \cite{Biscans2019}.
These small passive piezoelectric devices are bonded directly to the barrels of the test masses,
reducing the Q-factor of test mass mechanical modes and lowering the parametric gain below unity.
In previous observing runs, parametric instabilities required active damping using the test mass electrostatic drives \cite{PIESDdamping}.
With the addition of acoustic mode dampers the circulating power in the arm cavities has been increased by a factor of two.
The acoustic mode dampers increase the thermal noise contribution to DARM by less than 1\%.

Parametric instabilities have been observed at \SI{10.2}{kHz} and \SI{10.4}{kHz} at LHO during O3 \cite{Biscans2019}.
These frequencies are lower than the main target range of the acoustic mode dampers.
The instabilities were suppressed by tuning the end test mass ring heater (see Section~\ref{ss:tcs}) to shift the arm cavity higher-order-mode spacing away from the test mass acoustic mode.
Modeling of the arm cavity suggests that the overlap between the optical mode and acoustic mode is exacerbated by beam mis-centering on the test mass.
The beam is deliberately off-center to avoid known absorption features on the corresponding input test mass, as discussed in Section~\ref{ss:absorption}.
 
    \subsubsection{Radiation pressure torque}
    \label{ss:radiationpressuretorque}
    As the power circulating in the arm cavities increases, torques exerted on the test masses due to radiation pressure also increase.
These torques can produce instability when their magnitude approaches the restoring torque of the pendulum \cite{SidlesSigg_2006, Hirose_2010, Dooley_2013}.
While O3 power levels are still far from producing this instability condition, the torque modifies the dynamics of the suspended mirrors significantly and couples the angular motion of the cavity mirrors.
This requires angular control compensation filters to be modified as the optical power in the arm cavities increases.
This is discussed in Section~\ref{ss:angular}.
 
\subsection{Squeezer}
\label{ss:squeezer}
For O3 an in-vacuum squeezer was installed at each site to inject squeezed vacuum into the interferometers and reduce shot noise.
A full description of the new squeezer can be found in \cite{Tse2019}.
In contrast to previous squeezers for gravitational-wave detection \cite{Aasi2013,TheLIGOScientificCollaboration2011,Grote2013}, the squeezed vacuum source (an optical parametric oscillator) is placed inside the vacuum envelope on a separate suspended platform~\cite{Galiana2019}. This reduces squeezing ellipse phase noise and backscattered light noise \cite{Oelker2016a}.
The squeezer has been fully integrated into the automated lock acquisition sequence.

While Section~\ref{ss:power} discussed increasing the input power to the interferometer, which increases interferometer sensitivity by enhancing the gravitational-wave signal,
injecting squeezed vacuum improves the signal-to-noise ratio by decreasing the interferometer noise.
In this sense $\sim$\SI{3}{dB} of squeezing is equivalent to doubling the intracavity power to $\sim$\SI{450}{kW}.
The detector sensitivity is therefore closer to the Advanced LIGO design sensitivity, which specified \SI{750}{kW} intracavity power and did not include squeezing.

Above \sqzfreq{} the interferometer sensitivity is increased by \SI{\typicalSQZdBLHO{}}{dB} and \SI{\typicalSQZdBwithouterr{}}{dB} at LHO and LLO, respectively.
This provides a \typicalSQZBNSrangeincreaseLHO{} and \typicalSQZBNSrangeincrease{} increase in binary neutron star inspiral range at each respective site.

Below \qrpnfreq{}, injecting frequency-independent squeezed vacuum, as is done during O3, increases the quantum radiation pressure noise.
The low-frequency noise at LLO is small enough that this increase in quantum radiation pressure noise is detrimental to sensitivity and binary neutron star inspiral range.
The current squeezing level at LLO cannot be further increased without causing a reduction in range \cite{Tse2019}.
The squeezing angle is therefore set to $\sqzangdetuning{}^\circ$ from the optimal high-frequency configuration.  This increases range by reducing low-frequency radiation pressure noise at the expense of a \SI{0.5}{dB} increase in shot noise at high frequencies.  This effect is more fully explored in~\cite{Yu2020}.

Detuning of the signal-recycling cavity also produces frequency-dependent squeezing.
This effect was used to identify and correct a 2--\SI{3}{nm} detuning in the signal-recycling cavity length locking point at LLO.
Signal-recycling cavity detuning was then also exploited to maximize binary neutron star inspiral range.
 
\subsection{Core optic replacement}
\label{ss:coreoptics}
Several of the core optics were replaced before O3 to improve detector sensitivity, stability, and lock acquisition performance. The motivation and performance benefit of each replacement is presented here. 

At both sites the two end test masses were replaced.
To improve the lock acquisition sequence, the optical coatings on the O3 end test masses have lower scatter loss and increased reflectivity for \SI{532}{nm} laser light.
This increased the green arm cavity finesse from 15 to 70 at LHO and to 100 at LLO (more values in Appendix~\ref{s:table}).
This improves the reliability of the early stages of lock acquisition, where control of each arm length is transitioned from green to infrared error signals \cite{Staley:thesis}.

The $\sim$\SI{10}{ppm} reduction in scatter loss has resulted in improved power-recycling gain at both sites.  However when increasing the circulating power in the arm cavities, the power-recycling gain has not increased as expected due to nonuniform absorption on the optics increasing scatter losses in the arm cavities; see Section~\ref{ss:absorption}.

The X-arm input test mass at LHO was replaced before O3 following the identification of a point absorber in the coating.
The presence of the point absorber limited high-power operation and coupled jitter noise from the pre-stabilized laser to DARM.
The new input test mass shows no significant absorbers.
Similar defects have been found on several other test masses currently installed; these are further discussed in Section~\ref{ss:absorption}.

The signal-recycling mirror (SRM) at both sites was replaced.
The previous SRM was an aluminum and fused-silica composite with a 2" diameter optic that allowed for easy mirror replacement.
The composite SRM introduced thermal noise due to internal modes of the composite system with high mechanical loss.
The replacement SRM is monolithic fused silica, \SI{150}{mm} diameter, with no measurable thermal noise contribution to DARM.
To maximize the binary neutron star inspiral range, the SRM transmission should be reduced with increasing circulating optical power.
For O3, the SRM transmission was reduced from 37\% to 32\%.

The reaction masses, which are suspended in a separate pendulum chain behind the end test masses, provide high-frequency actuation via the electrostatic drive \cite{Aston2012}.
The proximity of the reaction mass to the end test masses can increase the damping noise due to residual gas bouncing between the test mass and reaction mass.
This noise is known as squeezed film damping \cite{PhysRevD.84.063007}.
Before O3 the reaction masses were replaced with annular reaction masses with cored out centers that retained the original electrode pattern.
These annular end reaction masses are expected to have reduced the squeezed film damping noise by a factor of 2.5 below \SI{100}{Hz} \cite{squeezed_film_montecarlo}.
 
\subsection{Test mass discharge}
\label{ss:discharge}
Charge that builds up on the test masses or changes in charge distribution around the test masses can result in electric field noise coupling to test mass motion, as discussed in Section~\ref{ss:charging}.
Several changes have been made to reduce charge build-up, discharge or depolarize the test masses, and monitor sources of electric field noise.

A likely source of test mass charge in O1 and O2 was ionization from UV light emanating from the ion pumps that are part of the vacuum system.
These pumps were relocated and baffled to prevent incident UV radiation on the test mass.
The electrostatic drive applies a large bias voltage $\sim\SI{100}{V}$ between electrodes.
Even with the aforementioned changes to ion pumps the effective charge on the test masses (witnessed by the electrostatic drive actuation force) changes slowly over time due to charge migration resulting in polarization \cite{Hewitson_2007,Prokhorov2010}.
This change in effective charge has predictable rate and direction.
Therefore the polarity of the electrostatic drive bias voltage is reversed periodically, reversing the direction of charge build up, thereby limiting the effective charge on the test mass.

The removal of polymer First Contact, which is used to clean test mass surfaces before a vacuum chamber is closed, results in triboelectric charging on the test mass surface.
A discharge procedure has been developed where ionized dry nitrogen gas is used to discharge the optic after the removal of First Contact \cite{Campsie_2011}.
Additionally, a test mass discharge system has been installed and demonstrated to effectively discharge with optics without opening the vacuum tanks.
This system again uses ionized nitrogen to flood the chamber up to \SI{30}{torr}.
 This results in up to an order of magnitude reduction in charge as interrogated by the electrostatic actuation force on the test mass.  

Finally, electric field meters were installed in the chamber of the Y end test mass of LLO and the X end test mass of LHO.
This electric field meter is designed to witness any time-varying electric field in the chamber that could induce a large enough force on a charged test mass to impact interferometer sensitivity.
 
\subsection{Stray light control}
\label{ss:scatterUpgrade}
Light that scatters out of the main interferometer beam can pick up time-varying phase relative to the main beam by reflecting or scattering again off moving surfaces.
If this scattered light is re-coupled to the main beam, it can produce noise in DARM, as discussed in Section~\ref{ss:scatterNoise}.

Observations of stray light in video monitors at LLO, and observations of anthropogenic ground motion coupling to DARM at LHO, led to further stray light investigations.
During O2 improved dumping of ``ghost" beams and damping of resonances of a stray-light baffle reduced scatter noise in DARM.
Based on this experience, multiple baffles were added or modified between O2 and O3.

Scattering mitigation activities continued during the break in O3, including additional baffle installation.
During the first half of O3, fringe wrapping noise was often observed at LLO when ground motion below \SI{10}{Hz} was large.
Drops in binary neutron star inspiral range visible in Figure~\ref{fig:range} were a daily feature at LLO.  Baffles installed around a transmitted light monitor at an end station at LLO reduced scattering noise from anthropogenic vibrations during the day.

An in-vacuum window traversed by the output beam was removed at LHO to mitigate scattered light coupling.
During O3, accelerometer tests at LHO localized a scattering site causing a \SI{48}{Hz} peak in DARM to a particular vacuum chamber.
A beam dump installed outside a viewport to the chamber dumped a stray beam and significantly reduced this noise contribution.
Late in the O3 run, transients from scattered light were greatly reduced at both sites by actuating on the reaction chain to reduce micron-scale relative motion between test masses and reaction masses.
The likely scattering path involved multiple reflections between the gold traces of the electrostatic drive on the reaction mass and the high-reflectivity coating on the test mass~\cite{Soni2020}.

\subsection{Alignment sensing and control}
\label{ss:angular}
In this section upgrades to the alignment sensing and control (ASC) scheme are reviewed.
Recent upgrades accommodate increased optical power, reduce noise injection to DARM, and produce more robust angular control.

These modifications to the alignment sensing and control scheme have produced the most robust angular noise control to date.
However, as discussed in Section~\ref{ss:ASC}, the overall noise contribution from angular controls is still the most significant source of noise in DARM below \SI{20}{Hz}.
Ongoing research aims to reduce angular control noise further~\cite{Mow_Lowry_2019}.

    \subsubsection{Radiation pressure compensation}
    \label{ss:radiationpressurecompentation}
    The radiation pressure dynamically links the angular motion of the test masses of each arm together via the mis-centering of the beam spots.
The angular modes of the cavity thereby created are called ``hard'' and ``soft'' modes~\cite{Barsotti2010}, since they respectively increase or decrease the stability of the resonant cavity.
The changes in system dynamics due to radiation pressure on the test masses require control filters to be modified as the power is increased.

The hard mode ASC control loop bandwidths must be sufficiently large to suppress arm cavity motion to maintain stable operation.
The unity gain of these loops is currently set at \SI{3}{Hz}, above the resonance of the Sigg-Sidles hard modes \cite{SidlesSigg_2006} at the circulating power.
Each site has adopted a different approach for the control of these degrees of freedom.

At LLO, hard mode filters optimized for plant dynamics up to \SI{25}{W} incident on the power-recycling mirror are used for the initial part of the lock acquisition sequence.
Just after reaching full input power, the control filters are switched to control filters optimized for \SI{40}{W} operation.
Cutoff filters, carefully tuned to minimize noise when used in conjunction with the high-power angular control compensation filters, are then engaged in the final stages of the lock acquisition sequence.

At LHO, an adaptive hard mode filter design has been implemented that allows continuous variation of the control filters for a range of input powers \cite{Yu:thesis}.
The filter is designed to correct for the radiation pressure torque effects, returning the plant dynamics back to that of a lower circulating power, chosen to correspond to \SI{10}{W} input power to account for uncertainties in the compensation. There is then only one control filter design for all power levels.

The soft mode controls mainly damp an instability around \SI{0.5}{Hz} using the quadrant photodetectors in transmission of the arms as the error signal.
The \SI{0.5}{Hz} pitch oscillation can be explained by a spurious dependence of the circulating power on the beam spot position that in turns creates an additional torque through the length-to-pitch cross-coupling of the suspensions. Controlling this instability has been a challenge towards operation at high power.
 
    \subsubsection{Alignment dither system}
    \label{ss:alignmentdithersystem}
    Changes in spot positions on core optics between locks have been shown to significantly alter the angular optical-mechanical plant.
To make the locking process more consistent, both sites have adopted a dither alignment system. 

Dithering is an intentional angular injection into an optic at a specific frequency.
Angular modulation is injected by actuation on the core optic penultimate masses.
For LLO the dithering frequencies are chosen outside of the detection band, below 10 Hz, whereas they are between \SIrange{15}{20}{Hz} for LHO.
The relevant length motion signal is demodulated at the same frequency to produce a measure of the angle-to-length coupling.
This signal is minimized by slowly adjusting the optic angle so that the rotation point coincides with the beam position.

The preferred beam position on each test mass is first determined in terms of minimum power losses in the cavities.
Each optics rotation point is digitally set to match these positions.
The error signals are then used to control the spot positions on the input and end test masses with bandwidths on the order of \SI{0.01}{Hz}.
This alignment dither system provides more repeatable spot positioning compared to quadrant photodetector error signals.
Being a reliable way to scan the surface of the mirrors for minimum losses, it also allowed the precise beam locations to be chosen to avoid point absorbers (Section~\ref{ss:absorption}).
 
    \subsubsection{Signal-recycling cavity alignment}
    \label{ss:signalrecyclingcavityalignment}
    Before O3, the angular control of the signal-recycling cavity proved challenging due to the lack of a good sensor.
Previously, the alignment error signal of the signal-recycling cavity was formed from the beat note between the \SI{9}{MHz} and the \SI{45}{MHz} sidebands.
This \SI{36}{MHz} beat note was detected by wavefront sensors at the antisymmetric port,
using the \SI{9}{MHz} \tem{00} as the reference beam and \SI{45}{MHz} \tem{01} and \tem{10} modes as the misalignment signals.

This signal-recycling alignment error signal was problematic.  The signal is weak due to inefficient transmission of the \SI{9}{MHz} sideband to the dark port.
Because the beat note is formed from two sidebands, the \SI{36}{MHz} error signal is not zero when the cavity is well aligned.  This results in some degeneracy with the beam centering on the wavefront sensor.
Additionally, higher-order modes generated by thermal distortions in the test masses can produce competing beat note signals that dramatically change the  \SI{36}{MHz} error signal response to signal-recycling cavity misalignment.

To generate a cleaner signal-recycling alignment error signal a new RF phase modulation sideband was injected into the interferometer at \SI{118}{MHz}, the 13th harmonic of the \SI{9}{MHz} signal.
The new alignment error signal is derived from the beat note at \SI{72}{MHz} between the \SI{118}{MHz} \tem{00} transverse mode and the \SI{45}{MHz} \tem{01/01} transverse modes.
The \SI{118}{MHz} sideband is more efficiently transmitted to the dark port than the \SI{9}{MHz} sideband due to the Schnupp asymmetry.
While the \SI{118}{MHz} sideband also suffers thermal distortions and \SI{72}{MHz} is also formed from two sidebands, these effects are manageable with a stronger signal.
The \SI{72}{MHz} beat note provides a robust signal for signal-recycling alignment control \cite{Yu:thesis}. 
\subsection{Lock acquisition and stability}
\label{ss:lockAqStability}
This section highlights instrument improvements that affect detector duty cycle.
Sections~\ref{ss:seismicupgrade}, \ref{ss:3_3HzInstability}, and \ref{ss:dac} provide examples of stability improvements.
Section~\ref{sss:carmoffsetreduction} is an example that makes the lock acquisition sequence more robust and hence faster.  
This compensates additional features, such as squeezing, that have made lock acquisition slower.
The result is a minimal change in average lock acquisition time.
Section~\ref{ss:vco} is an example of the mitigation of large transients that can make data unusable.  
The resulting improvement in the detector duty cycle was discussed in Section~\ref{ss:dutycycle}.

    \subsubsection{Seismic controls}
    \label{ss:seismicupgrade}

Core and auxiliary optic suspensions are mounted on isolation platforms that serve to decouple the optic motion from the ground.
Different types of platform are used for the core and auxiliary optics, but the general concept is identical for all of them.
Each platform provides a combination of passive and active isolation to bring the platform motion down to $\sim$\SI{10e-11}{m/\rtHz} at \SI{10}{Hz} \cite{Matichard_2015}.
Between O1 and O3, hardware and software changes have improved the seismic configurations at both sites, with improved lock stability in the presence of increased ground motion.

Beam rotation sensors, which measure ground tilt, were installed at both end stations (LHO) and the end and corner stations (LLO); these facilitate tilt-corrected ground motion measurements \cite{Venkateswara_2017,windproofing}.
This is relevant below \SI{0.1}{Hz} where seismometer signal is contaminated by ground tilt, which is exacerbated under windy conditions.
In the corner station at LHO, tilt-free ground motion is measured by a seismometer at the center of the building, where ground tilt is reduced compared to the edges of the building.
Effective cancellation of ground motion between 0.1 and \SI{1}{Hz} is performed by feedforward from seismometers \cite{Matichard_2015}.
This can inject excess noise at out-of-band frequencies.
With tilt removed, a more aggressive control configuration is possible, allowing the interferometers to have better resilience to windy or high microseism conditions.
Tilt cancellation is the main upgrade to the seismic system between O1 and O2 for LHO, and O2 and O3 for LLO.

Elevated ground motion in the 0.03 to \SI{0.1}{Hz} band during an earthquake can overwhelm seismic isolation platforms and unlock the interferometer.
During an earthquake the end stations and corner station predominantly experience common ground motion.
For O3 an experimental sensor configuration was implemented upon an early-warning earthquake trigger, calculating the common motion by averaging seismometer signals from the corner and end stations, and then subtracting this from local ground sensors to produce feedforward signals \cite{Lantz2018, Coughlin2017, Biscans_2018,schwartz2020improving}.
This system has improved the detector duty cycle by allowing the interferometers to remain locked through moderate earthquakes.

\subsubsection{Suspension chain damping}
    \label{ss:3_3HzInstability}
    The test masses are suspended by a main quadruple suspension chain.
Each of these main chains is accompanied by a quadruple stage reaction chain, which sits behind the test masses.
The suspension chains are designed to have low mechanical loss to minimize thermal noise, hence the resonant modes of the suspension chain have high quality factors.
Sensing and actuation for local damping of these modes occurs at the top mass of each chain \cite{strain2012damping}.
This limits the injection of local sensing noise into the gravitational-wave readout through three stages of mechanical filtering at the cost of reduced actuation authority at the test mass.

Interferometric control signals, which are more sensitive than local damping signals, are applied as actuation between the main and reaction chains.
These control signals can introduce energy into the modes of both suspension chains.
While the fundamental vertical (bounce) and roll modes of the test masses were actively damped in O1 and O2, these are now passively damped with tuned mass dampers \cite{RobertsonBRD2017}.
Many of the fundamental, second, and third harmonics of the transverse (``violin") modes of the fused silica fibers supporting the test masses were actively damped at LLO and LHO during O3.
In the future these modes may be passively damped as significant time is lost when they are excited either through control system failure or excitation from earthquakes. During O3, the amplitude of a pitch mode of the input test mass reaction chain became unstable at LLO, producing excess motion at \SI{3.3}{Hz}.
To address this issue, local signals from the penultimate stage of the suspensions are filtered and applied to the reaction chain's top mass, enhancing energy extraction.
 
    \subsubsection{CARM offset reduction}
    \label{sss:carmoffsetreduction}
    Reducing the common arm length (CARM) offset to bring the arms onto resonance with the carrier field has historically been a fragile point of the LLO lock acquisition sequence.
This was due to a handoff between a CARM error signal derived from the carrier field transmitted from the arm cavities to an error signal derived from the \SI{9}{MHz} field reflected from the power-recycling cavity.
This handoff was historically performed with the common arm lengths detuned \SI{10}{pm} from resonance.
Near this point, the gain of the \SI{9}{MHz} signal passes through zero~\cite{MartynovPhD}. Non-ideal conditions during this handoff would result in the wrong sign control gain, departure from the fringe, and lock loss.
In O3, the transmission lock is left engaged while engaging the \SI{9}{MHz} lock, then a reduction in CARM offset is applied just before ramping down the transmission loop gain.
This nudge pushes the \SI{9}{MHz} signal into the regime where the \SI{9}{MHz} gain is increasing, ensuring it reaches the correct set point and significantly increasing the CARM offset reduction stability.
 
    \subsubsection{New VCO at Livingston}
    \label{ss:vco}
    The LIGO detector characterization group identified a type of loud transient signal (``glitch'') known as a whistle glitch~\cite{MarissaPhD}.
The glitch morphology is a narrow spectral line rapidly moving in frequency that often appears to ``reflect'' off zero.
Whistle glitches are thought to be the result of radio frequency lines crossing each other, beating to produce a signal in the gravitational-wave measurement band.
This type of glitch is particularly problematic as the ``whistle" can produce signals similar to those produced by coalescing astrophysical binaries.
Investigations at LLO identified a voltage controlled oscillator (VCO) that is at least in part responsible for producing whistle glitches.  This VCO produces an offset frequency between the laser light going to a reference cavity that is used for laser frequency noise suppression and the light going to the interferometer.
The VCO was replaced at LLO prior to the run and ongoing investigations during the run resulted in several changes.  The change in VCO resulted in a dramatic reduction in the occurrence of whistle glitches.
However some interim solutions suffered from increased laser frequency noise. Frequency noise represented in Figure~\ref{fig:LloNb} is a later iteration with less impact on high-frequency sensitivity.
Even in earlier iterations frequency noise had minimal impact on binary neutron star inspiral range.

At LHO whistle glitches appear when this VCO crosses certain frequencies.
No spectral features could be found at these frequency crossing points so the source of the second frequency line is unknown.
The VCO frequency is therefore chosen to avoid these crossing points.
However during periods of high ground motion, frequency excursions can become large resulting in whistle glitches.
Whistle glitches contaminate roughly 1\% of the data from LHO.  This was similar to LLO during the first two observing runs.
 
    \subsubsection{Increased actuator range}
    \label{ss:dac}

A number of changes have been made to the test mass actuators that improve detector duty cycle while limiting injected noise.

Before O3 all suspension actuators used 18-bit DACs (General Standards 18AO8) that suffer from a \emph{zero-crossing} issue.
When the digital signal crosses zero counts the output voltage may contain an impulse that can get worse in time without periodic re-calibration of the DAC.
This impulse injects broadband noise.
Additionally, the nominal noise has been found to depend on drive amplitude, DC offset, and the DAC channel used.
The actuators on the lower stages have been partially replaced with 20-bit DACs (General Standards 20AO8C500K), which have slightly lower noise and do not have this zero-crossing issue.

To better understand DAC noise, \emph{in situ} monitors were partially installed to measure the noise while removing the large optic control signal at low frequencies.
This allows for sensitive real-time monitoring of the control signal noise in the gravitational-wave band without saturation of the monitor due to low-frequency control signals.

The test-mass actuator analog electronic filter was also modified to provide greater actuation range above \SI{15}{Hz} without injecting significant noise.
Preliminary results suggest this modification has improved detector resilience against fast large transients that can unlock the interferometer.

These improvements have helped to move the actuator noise contribution safely below the current sensitivity and have produced more robust interferometer control.
  
\section{Interferometer characterization}\label{s:commissioning}
The noise budget presented in Section~\ref{s:noise} is one tool used to present and understand the interferometer noise floor.
In parallel, commissioning activities seek to characterize and optimize the detector, and search for new noise sources.
This section details some of the commissioning investigations which occurred before and during O3.
Section~\ref{ss:crosscor} discusses a cross-correlation measurement which reveals noise below shot noise,
Section~\ref{ss:ArmPowerMeas} explains a new method of measuring the circulating power in the interferometer arms,
Sections~\ref{ss:tcs} and \ref{ss:absorption} discuss the thermal compensation system and nonuniform coating absorption.
Section~\ref{ss:RFOscillator} discusses an investigation into radio frequency oscillator noise,
and Section~\ref{ss:charging} discusses test mass charging and stray electric fields.
\subsection{Correlated noise}
\label{ss:crosscor}
In DC-readout operation, the two DC photodetectors (DCPDs) at the antisymmetric port are summed together to measure the DARM degree of freedom.
By cross correlating the DCPD signals---and compensating for the additional correlation induced by the DARM control system---the sensing noise is averaged out and the correlated noise beneath can be revealed \cite{IzumiXcorr2017}.
Note that both shot noise and photodetector dark noise are averaged out.
This measurement is taken without squeezed light injection because this induces correlated quantum shot noise between the DCPDs \cite{McCullerCrossCorr2018}.

\begin{figure}[t]
    \subfloat[LHO]{
    \includegraphics[width=\columnwidth]{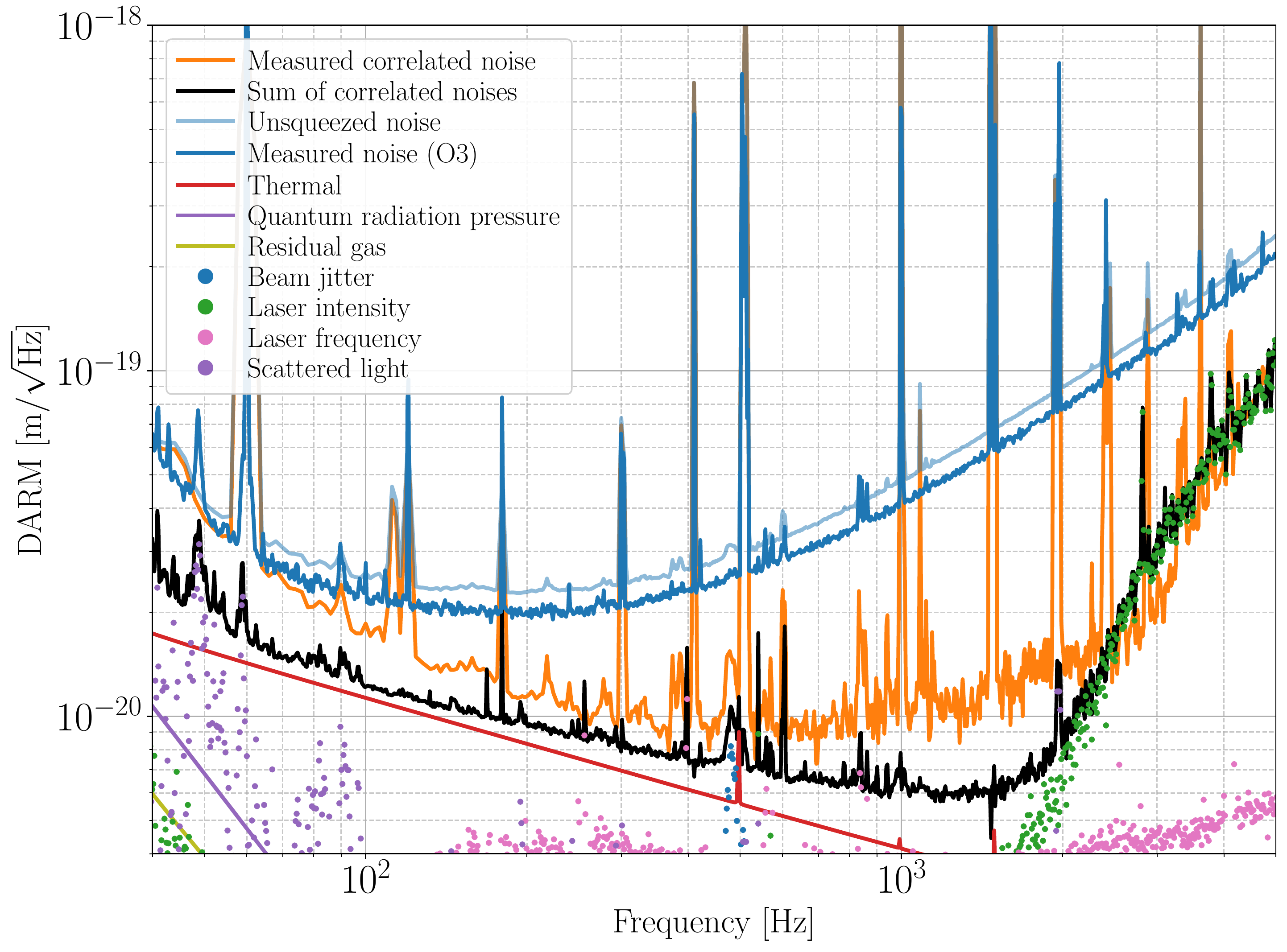}
    \label{fig:Lhoxcor}
  }\hfill
  \subfloat[LLO]{
    \includegraphics[width=\columnwidth]{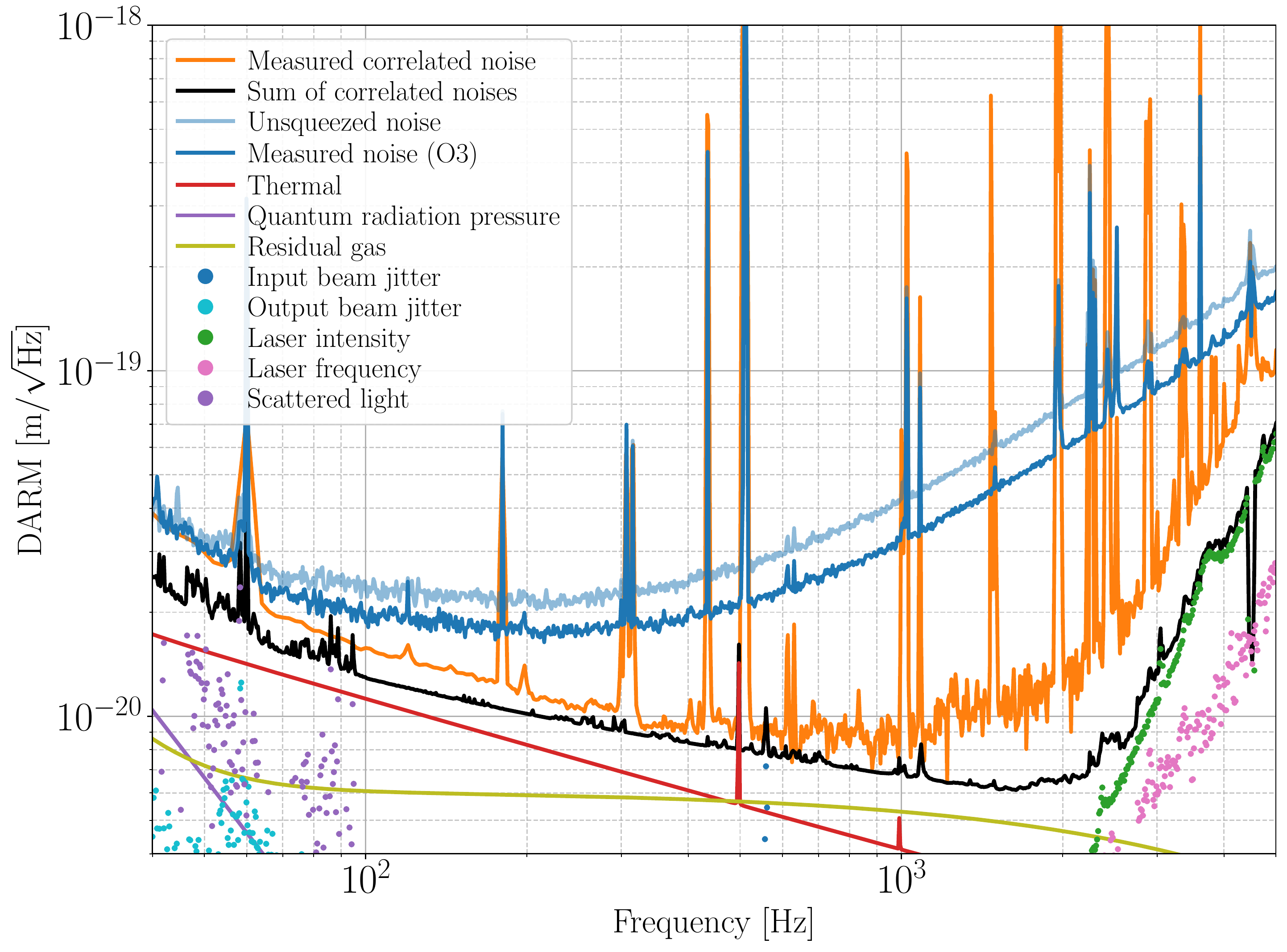}
    \centering
    \label{fig:Lloxcor}
  }
  \caption{
  Correlated displacement noise at the antisymmetric port, measured as described in Section~\ref{ss:crosscor}.
  The total expected correlated noise is shown in black.
  The dominant expected contribution to the correlated noise curve is thermal noise from \SIrange{60}{900}{Hz}.
  The upper two traces show the DARM noise around the time of the correlation measurement with and without squeezed light injection.
  The differences in beam jitter coupling is covered in Section~\ref{ss:jitter}.
  To average away uncorrelated sensing noise between the DCPDs, 
  the LHO measured correlated noise trace used \num{20000} averages over \num{10000} seconds of data, 
  while LLO used \num{21600} averages over \num{10800} seconds of data.
  }
  \label{fig:xcor}
\end{figure}

Figure~\ref{fig:xcor} shows the results of this measurement.
Sufficient averages were taken to reach below statistical noise across the entire range.

The mirror coating thermal noise is the limiting correlated noise source around \SI{200}{Hz}.
This represents a fundamental limit to the sensitivity improvement we can expect from lowering the quantum shot noise via increased squeezing and increased laser power.

This measurement provides some confirmation of the increased gas noise contribution at LLO.  
Without the gas noise contribution, the expected correlated noise between \SI{400}{Hz} and \SI{1}{kHz} deviates significantly from the measurement. 
The excess gas noise at LLO is the result of a leak around the mid point of the X arm vacuum enclosure.  
The leak has been fixed since the time of this measurement, as explained in Section~\ref{ss:gasnoise}.

At frequencies above \SI{3}{kHz}, laser frequency and intensity noise dominate the correlated noise budget.  
The poor match between predicted and measured correlated noise above \SI{2}{kHz} at LLO is thought to be due to the coupling function of laser (frequency and intensity) noise to DARM being non-stationary.

The origin of the correlated noise at \SIrange{1}{2}{kHz} is unknown, but is a factor of five below DARM sensitivity.
The origin of the correlated noise below \SI{100}{Hz} is also unknown, and is likely the same noise that limits DARM sensitivity at \SI{40}{Hz} in Figure~\ref{fig:nb}.

\subsection{Measuring the arm power}
\label{ss:ArmPowerMeas}
The circulating laser power in the arm cavities governs the amplitude of the interferometer response to gravitational-wave signals.
The arm power is difficult to estimate precisely due to large uncertainty in the power on the beamsplitter and optical gain of the arm cavities.  
Uncertainties are dominated by photodetector calibration and interferometer optical loss uncertainty.

The arm powers in a power-recycled interferometer with a 50:50 beamsplitter should follow
\begin{equation}
  \label{eq:armpower}
  P_\mathrm{arm} = \frac{1}{2} P_\mathrm{in} G_\mathrm{PR} G_\mathrm{arm},
\end{equation}
where $P_\mathrm{arm}$ is the power in an arm,
$P_\mathrm{in}$ is the input power,
$G_\mathrm{PR}$ is the power-recycling gain, and
$G_\mathrm{arm}$ is the arm power gain.

The input power $P_\mathrm{in}$ is the power incident on the power-recycling mirror, and is estimated from a pick-off just before entering the interferometer.
The power on the beamsplitter $P_\mathrm{BS}$ is estimated directly from a pick-off of the power-recycling cavity.
The power-recycling gain is estimated from the ratio of the power incident on the beamsplitter over the input power:
$G_\mathrm{PR} = P_\mathrm{BS} / P_\mathrm{in}$.
Finally, the arm power gain $G_\mathrm{arm}$ is estimated from the input and end mirror transmissions, as well as the round-trip loss.

Photodetector power uncertainty originates from uncertainty in calibration, losses along beam path combined with beam size mismatch and misalignment.  We have assumed a total uncertainty of 5\% in power estimated from pick-off photodetectors, $P_\mathrm{in}$ and $P_\mathrm{BS}$.
The arm gain $G_\mathrm{arm}$ at Livingston is assumed to be $265$ with uncertainty of 5\%.
The Hanford X-arm gain is $262$, while the Y-arm gain is $276$; the ~5\% gain difference is due to the slightly different transmissions of the input test masses at Hanford (see Appendix \ref{table:aLIGOParams}).
Results are shown in Table \ref{table:ArmPowers}.

A technique to measure the arm powers using radiation pressure was developed prior to O3 \cite{ArmPowerMeas2019, Izumi_FreqRespPart3_2015}.
The length of the signal-recycling cavity is modulated, creating audio sidebands on the carrier laser in the signal-recycling cavity.
The audio sidebands enter the arm cavities producing a light power modulation that has opposite sign in each arm cavity, causing a signal to appear in DARM.

The power estimate is derived from the ratio of measurements of the relative intensity noise in transmission of the arms and the DARM signal.
Many complexities, such as the amplitude of the signal-recycling sideband, photodetector calibration and optical losses between interferometer and photodetector, appear in the numerator and denominator and divide out in this measurement.
For frequencies below \SI{100}{Hz} radiation pressure moving the test masses dominates the DARM signal, and the transfer function between arm transmitted power relative intensity and DARM has a simple expression:
\begin{equation}
\frac{L_\mathrm{DARM}}{\mathrm{RIN}_\mathrm{arm}}(f) = \frac{2 P_\mathrm{arm}}{ m c \pi^2 f^2} = \frac{\alpha}{f^2},
\end{equation}
where $L_\mathrm{DARM} = L_x - L_y$ is the differential arm displacement,
$\mathrm{RIN}_\mathrm{arm}$ is the relative intensity of the arm transmission,
$m$ is the mass of the final stage of the quadruple pendulum, and
$P_\mathrm{arm}$ is the power in the arm.

By fitting the $\alpha/f^2$ slope of the relative intensity of the arm power transmission to DARM transfer function, the power in each arm can be estimated according to
\begin{equation}
  \label{eq:srcl_arm_power_meas}
  P_\textrm{arm} = \frac{1}{2} \alpha m c \pi^2.
\end{equation}

Each arm power estimate relies on the relative intensity response of quadrant photodetectors on transmitted beams from each arm.
Each quadrant's relative intensity response can be distorted by poor alignment, as small changes in alignment can result in one or more quadrants becoming saturated with light.
The spot positions on the arm cavity optics and transmission monitor table drift can affect the alignment onto the quadrant photodetectors.
These effects can bias the arm power estimate, and must be monitored to ensure the accuracy quoted in Table~\ref{table:ArmPowers}.

Table \ref{table:ArmPowers} reports the measured arm powers during O3.
Measurements derived from signal-recycling cavity length modulation are consistent and more precise compared with measurements derived from test masses' reflectivity (Table~\ref{s:table}) and beam power measurements.

\begin{table}
  \centering
  \begin{tabular}{ l c c c c }
    Power                                         & Symbol            & LHO            & LLO            & Units \\ \hline
    Input                                         & $P_\mathrm{in}$   & 34 $\pm$ 2     & 38 $\pm$ 2     & W     \\
    Power-Recycling Gain                          & $G_\mathrm{PR}$   & 44 $\pm$ 3     & 47 $\pm$ 3     & W/W   \\ X-arm via Eq. \ref{eq:armpower}               & $P_X$             & 190 $\pm$ 14   & 240 $\pm$ 18   & kW    \\
    X-arm via Eq. \ref{eq:srcl_arm_power_meas}    & $P_X$             & \lhoXarmPower{}& \lloXarmPower{}& kW    \\
    Y-arm via Eq. \ref{eq:armpower}               & $P_Y$             & 200 $\pm$ 15   & 240 $\pm$ 18   & kW    \\
    Y-arm via Eq. \ref{eq:srcl_arm_power_meas}    & $P_Y$             & \lhoYarmPower{}& \lloYarmPower{}& kW    \\

\end{tabular}

  \caption{
    Highest measured laser power levels during O3.
    Input power is estimated via a pick-off from the light incident on the power-recycling mirror.
    Power-recycling gain is estimated from the pick-off of the power-recycling cavity, using a ratio of power on the beamsplitter and input power.
    Arm powers are estimated in two ways.
    The first method is via input power and gain estimates, Eq.~\ref{eq:armpower}.
    Arm power uncertainties for Eq.~\ref{eq:armpower} are propagated from uncertainty in the input power, power-recycling gain, and loss in the arms.
    The second method is via radiation-pressure relative intensity noise to DARM transfer function, Eq.~\ref{eq:srcl_arm_power_meas}.
    Arm power uncertainties for Eq.~\ref{eq:srcl_arm_power_meas} are derived from the coherence of the measured transfer function.
    Typical arm power levels at LLO were about 5\% lower over the course of the run.
  }
  \label{table:ArmPowers}
\end{table}
 
\subsection{Thermal compensation}
\label{ss:tcs}
The thermal compensation system (TCS) is designed to measure and actuate on the thermal lenses and radii of curvature of the core optics \cite{Brooks:2016}.
The operational target for the TCS is to correct optical aberration induced by absorption in core optics and to correct any static lens discrepancies.

The core TCS actuators consist of ring heaters situated around the barrel of each test mass and CO$_2$ lasers which heat the compensation plate behind the input test masses (ITMs).
The ring heaters create a negative thermal lens in the test masses and reduce the radius of curvature on the high reflectivity surface, while the CO$_2$ lasers can create either a positive or negative lens in the compensation plate.
Other TCS actuators that have been tested include a disk heater behind ``SR3," a reflective optic in the signal-recycling cavity, to decrease its surface curvature,
and a CO$_2$ laser projected onto the signal-recycling mirror (SRM). The accumulated substrate thermal lens of each test mass is monitored using Hartmann wavefront sensors \cite{Brooks:07}.
The two sites use different TCS settings and TCS settings have changed during O3.

At LLO, only ITM ring heater actuators are used.
Applying a heat load to the ITM actuators in common is expected to affect optical build up in the coupled power-recycling cavity and also mode matching to the output mode cleaner.
Applying heat load differentially to each ITM actuator is expected to affect the contrast defect, resulting in changes in frequency and intensity noise coupling to DARM \cite{Brooks:2016} (Sections~\ref{ss:frequency} and \ref{ss:intensity}).
Even after minimizing noise coupling with the differential degree of freedom, it was found that the common degree of freedom could further decrease laser noise coupling.

Suboptimal laser noise coupling affected LLO binary neutron star inspiral range more than suboptimal power in the power-recycling cavity.
Therefore the ring heaters are tuned to minimize laser noise coupling.
Optimal mode matching of the interferometer beam to the output mode cleaner was inferred to occur at approximately the same ring heater setting as the minimum in laser noise coupling.
This inference was made based on measurements of the interferometer response to a mechanical excitation of a test mass as a function of ring heater power at LLO.
It was also found that the ring heater settings that minimized laser frequency noise coupling were not exactly the same as those required to minimize laser intensity noise coupling.
Therefore after the change in voltage controlled oscillator (see Section~\ref{ss:vco}), the dominant coupling switched from frequency to intensity noise.
The ITM ring heaters were tuned differentially to minimize intensity noise coupling.

No improvement in binary neutron star inspiral range, mode matching, or noise coupling could be achieved with the SR3 heater at LLO.
By applying positive and negative SRM thermal lens with the CO$_2$ laser, a limit was set on the amount an SRM lens could improve mode matching to the output mode cleaner of $<1\%$.

At LHO, increased absorption in the Y-arm ITM (ITMY) causes a power-dependent mismatch between the two arms.
To maintain stability while increasing the input power, the TCS is used to preheat the test masses while acquiring lock.
In this scheme, when the interferometer is acquiring lock with low input power, the compensation plate CO$_2$ lasers create a thermal lens that emulates the thermal lens due to absorption in the ITM coatings at operating power.
When power builds up in the arm cavities (Table~\ref{table:ArmPowers}) the CO$_2$ laser power is reduced such that the thermal transient in the ITMs roughly cancels the thermal transient in the compensation plates.
At full input power the ITM CO$_2$ and ring heater settings are tuned to minimize laser frequency noise coupling, the dominant laser noise coupling at LHO.
This ultimately resulted in an increase in ITMX's CO$_2$ laser power \cite{VoPhD}. 

The effect of the SR3 heater was also studied at LHO.
At LHO a \SI{4}{W} power setting was shown to improve many parameters including binary neutron star inspiral range.
As described in Section~\ref{ss:parametric}, at LHO the end test mass ring heaters were used to change arm cavity transverse mode spacing to avoid parametric instability.

Avoiding parametric instability is complicated by changes in mirror radii of curvature resulting from absorbed optical power in the mirror coating.
This changes the tuning condition for parametric instability and in O2 resulted in transient instabilities occurring in the first few hours of operation.
A scheme described in \cite{Hardwick2020_DTC} was demonstrated whereby transients applied to the ring heating null the transient in mirror radii of curvature.
This allowed the optical power to be increased to \SI{170}{kW}, demonstrating precise control of the cavity geometry.
While this scheme has not been used since the installation of acoustic mode dampers discussed in Section~\ref{ss:parametric}, at higher optical power it could be useful to reduce interferometer transients and quell remaining instabilities.

The thermal compensation tuning and noise couplings discussed above are complicated by the presence of point absorbers in the test mass coatings discussed in Section~\ref{ss:absorption}.
 \subsection{Nonuniform coating absorption}\label{ss:absorption}

Increasing optical power on the interferometer power-recycling mirror from \SI{25}{W} to $\sim$\SI{40}{W} did not result in a proportional increase in optical power in the power-recycling cavity.
The loss of optical buildup cannot be recovered with adjustments to the thermal compensation actuators described in Section~\ref{ss:tcs}.
When the input power is increased the power-recycling gain degrades with a time constant of $\sim$\SI{100}{s}, while the time constant of a uniform absorption thermal lens is $\sim$\SI{1000}{s}.
These features indicate nonuniform absorption resulting in increased optical loss.

\begin{figure}[hbt!]
  \centering
  \includegraphics[width=\columnwidth]{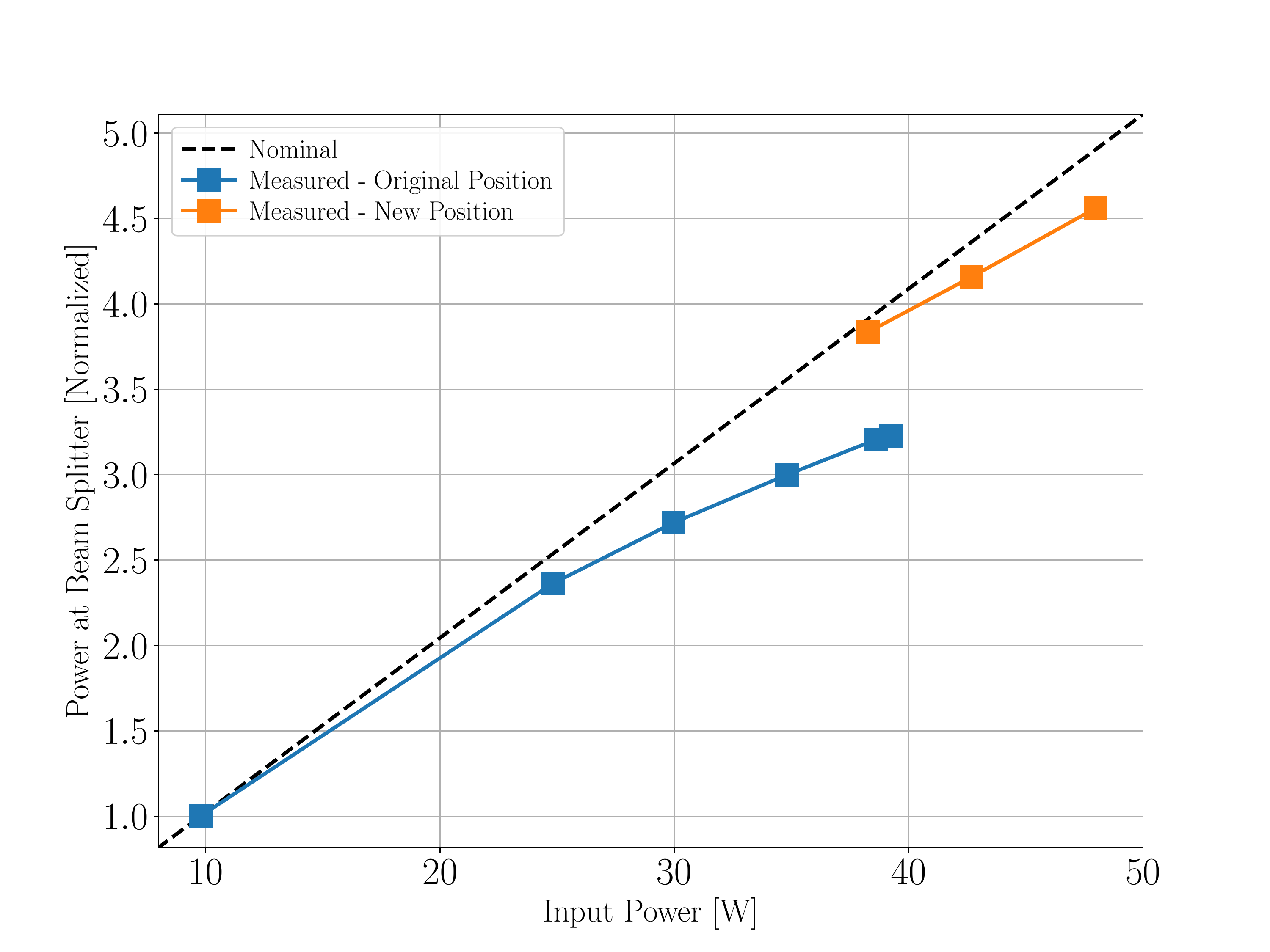} \caption{
  Power incident on the beam splitter as a function of interferometer input power at LLO.
  The dashed line is the expected power if losses are independent of power, while other traces are the measured power at the pick-off port.
  Point absorbers thermally distort the test masses and increase optical loss through scattering, resulting in the lower trace.
  Moving the beam spot on the end test masses to avoid these absorbers and maximize the recycling gain can partially mitigate these losses, leading to the upper measured trace.
  }
  \label{fig:point_loss}
\end{figure}

Moving the beam position on the test masses showed a position-dependent optical loss.
Figure~\ref{fig:point_loss} shows the measured relation between input power and recycling power with two traces for two different beam positions.
The difference between the expected and measured power at the beamsplitter could be reduced with adjustments to the beam spot position on the test masses,
moving the spot position away from an area of high absorption.

\begin{figure}[hbt!]
  \centering
  \includegraphics[width=0.7\columnwidth]{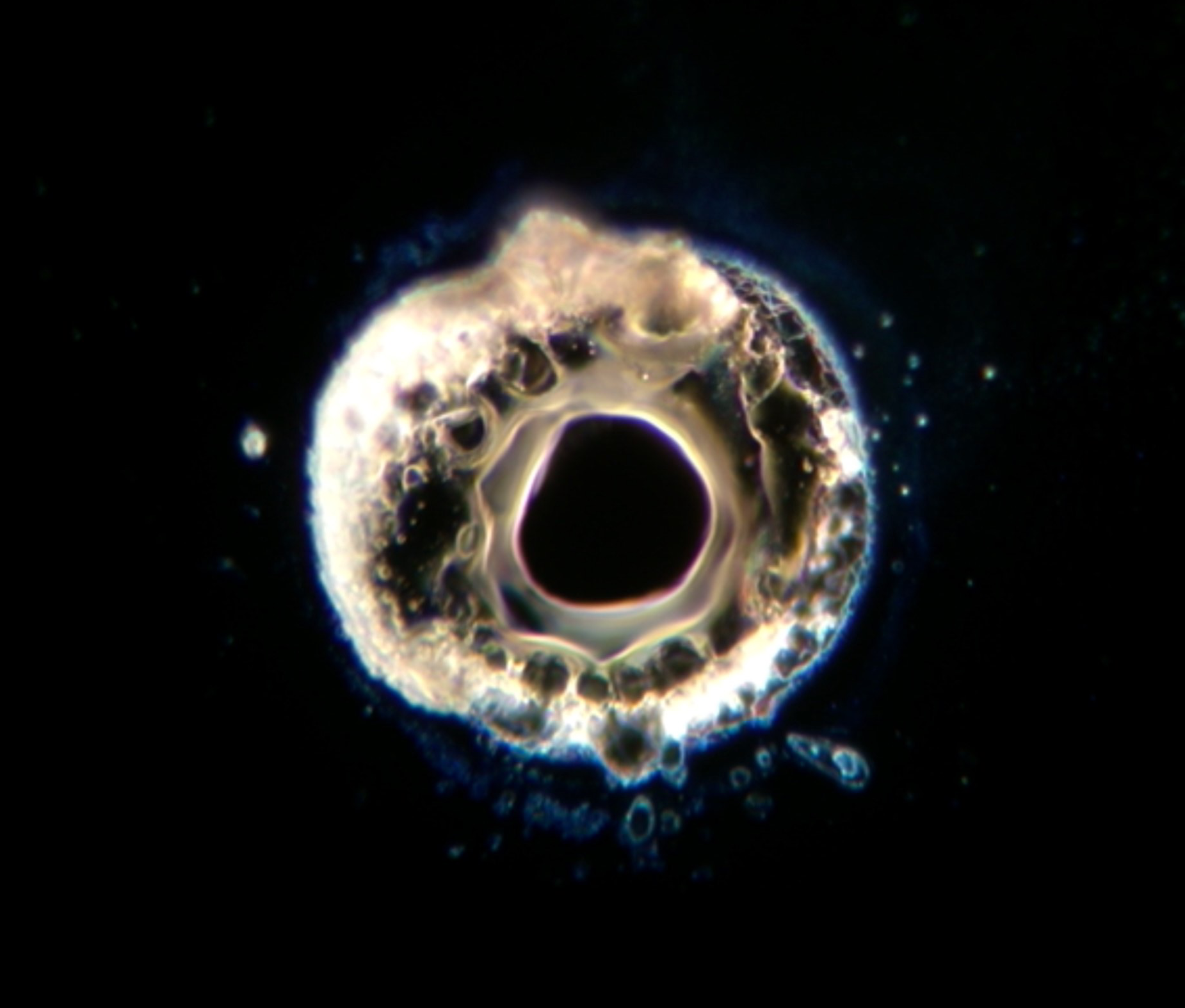}
  \includegraphics[width=0.8\columnwidth]{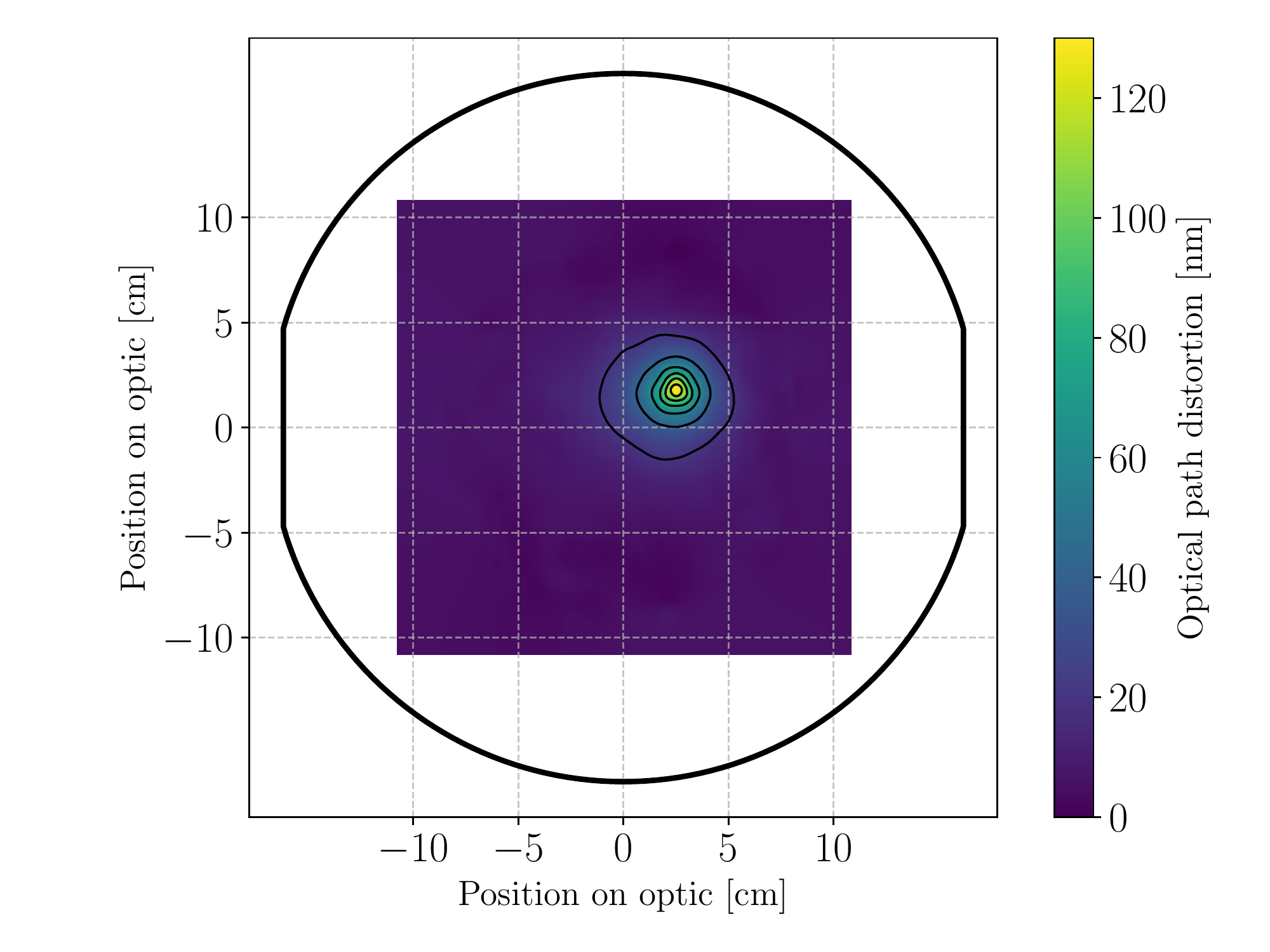}
  \caption{
  Point absorbers imaged using dark field microscopy (top) and \emph{in situ} with the Hartmann wavefront sensor (bottom).
  The top image shows the point absorber on LHO's ITMX for O1 and O2.
  The absorber is \SI{155}{\um} across the bright center.
  The bottom image is a Hartmann wavefront sensor image of the same optic in the interferometer. The main interferometer beam uniformly heats a region roughly \SI{5}{cm} in radius and illuminates the defect, causing a point distortion in the wavefront.
  There is an uncertainty of about \SI{\pm 1}{cm} in the location of the origin of the Hartmann wavefront sensor coordinate system.
  The largest contour ring represents a \SI{20}{nm} optical distortion and the contour spacing is \SI{20}{nm}.
  }
  \label{fig:hws_image2}
\end{figure}

The Hartmann wavefront sensors at both sites revealed nonuniform localized absorption or \textit{point absorbers} on several optics in O2 and O3.
In O3 they are present on LLO's ETMX and ETMY and LHO's ITMY and ETMX.
They have been unaffected by attempts to clean them from the optic surfaces.
Figure~\ref{fig:hws_image2} shows a microscope image and Hartmann wavefront sensor image of a point absorber which was present on LHO's ITMX during O2; this optic was replaced for O3.
The Hartmann wavefront sensor image is taken \emph{in situ}, while the microscope image is taken after removal from the interferometer.

Simulation and analysis presented in \cite{Brooks2020} broadly confirm the level of observed optical loss is that expected from point absorbers.  The mechanism for optical loss is thermo-elastic expansion distorting the test mass surface resulting in light being scattered from the fundamental cavity mode.

Microscopic analyses of mirror coatings on spare test masses are ongoing.
Features identified on test masses that have been removed from the interferometer have been compared to features identified on test masses that have not yet been installed.
These investigations suggest that the point absorbers are likely due to contamination introduced during the coating process and are therefore present before installation.
They may be altered by the intense laser power they experience in the arm cavity.
Efforts are ongoing to image potential absorbers using thermal cameras and Hartmann wavefront sensors before test mass installation.

\subsection{RF oscillator noise investigations}
\label{ss:RFOscillator}
Radio-frequency (RF) phase modulation sidebands are imprinted on the input beam to the interferometer to control the various longitudinal and angular degrees of freedom.
The electro-optic modulator which imprints the RF sidebands also imprints oscillator phase and amplitude noise on the input beam.
The RF oscillator noise is not directly sensed or controlled beyond the electro-optic modulator driver amplitude stabilization.
Because the RF sidebands are designed to not resonate in the full interferometer, the RF oscillator noise is not filtered by the common arm cavity pole.
Additionally, the RF sidebands do not carry gravitational-wave signal but do exit the antisymmetric port at higher overall power than the carrier in operation.
The output mode cleaner is placed at the antisymmetric port to lock onto the main carrier light and reject the RF sidebands.
The rejection is not perfect, and a small amount of RF sideband light imprints its noise on the DARM signal.
RF oscillator noise coupling was investigated in the first observing run, see Section III G of \cite{PhysRevD.93.112004}.

At LHO, a 9th-order transverse optical mode was visible on the output mode cleaner cavity transmission camera. Modeling of the output mode cleaner suggested that this mode on the upper \SI{9}{MHz} sideband could be close to co-resonant with the carrier given the cavity geometry. The output mode cleaner length is controlled via two piezoelectric transducers (PZTs) attached to the two curved mirrors.
The offset locking voltage of one of the PZTs was large, possibly flexing its curved mirror and changing the cavity geometry to allow this light to transmit through the cavity.
A new PZT driver was installed for the purpose of relieving the high offset voltage needed to lock the output mode cleaner.
Tests changing the PZT offset voltage between locks did not yield conclusive results in the differential arm noise.

\subsection{Optic charging and stray electric fields}
\label{ss:charging}

Length actuation on the test mass is performed using electrostatic drivers (ESDs)~\cite{Aston2012}.
Both \emph{in situ} and laboratory tests suggested charge separation due to a water monolayer on the optic~\cite{Awakuni1972,Ho1967} was producing a change in actuator strength of a few percent over weeks~\cite{Prokhorov2010, Buikema2020}.

A large earthquake in Montana in July 2017 decreased the sensitivity of LHO at low frequencies, hypothesized to be due to charging of a test mass by rubbing against an earthquake stop.
Additionally, excess coupling between motion of the protective cage surrounding the suspension and the test mass was observed; subsequent discharging of the optic by ionized gas significantly reduced this coupling and the noise.
In spite of this, large actuator bias voltages injected noise into DARM that suggested excess unexplained electric fields.

The steel vacuum chambers act as Faraday cages and largely shield the optics from external electric fields, but in-vacuum electronics and signals entering through uncovered viewports can still couple to DARM. As such, an electrometer was installed at each interferometer next to an end test mass to search for large-scale in-chamber electric fields. Measurements were consistent with the noise floor of the instrument at \SI{3}{\uV/m/\rtHz} at \SI{100}{Hz}. Viewport injections of electric fields confirmed the electrometer as a good witness, but the coupling to DARM was at least \num{100} times below the current DARM sensitivity. Large excitations of the suspension cage motion were not seen in DARM. Together these suggest that net optic charge is and remains low.

Local charge separation/polarization can still affect how local electric fields couple to optic motion. Measurements of the coupling of sources of these fields to DARM was performed to estimate the noise contribution for arbitrary actuator configurations~\cite{Buikema2020}. It was discovered that ground currents can produce voltage fluctuations of the ESD driver, producing a potential difference between the ESD and the cage that is not filtered by the driver electronics. This mechanism was likely the source of narrow spectral features that had previously been removed by partially isolating the ESD electronics \cite{Covas:2018}; additional reconfiguration of these electronics helped to eliminate other noise sources.

\section{Future Work}
\label{s:futurework}

Work continues on improving the sensitivity and duty cycle of the observatories.
O3 was paused for a one-month commissioning break during October 2019 to install several upgrades including in-vacuum baffles for mitigating scattered light noise (Section~\ref{ss:scatterNoise}).
Wind fences were installed at the Hanford end stations to reduce wind shear on the buildings and lab floor motion in windy conditions; this is expected to improve the network duty cycle.
An extended upgrade and commissioning period will precede observing run four (O4) \cite{Abbott2018}.

A goal for O4 is to improve the shot-noise-limited sensitivity, with increased intracavity laser power.
To this end, an additional free-space amplifier stage (neoVAN-4S-HP) outputting up to \SI{114}{W} will be installed \cite{Thies2019}.
The addition of the acoustic mode dampers make operation at higher powers possible with minimal parametric instability (Section~\ref{ss:parametric}) \cite{Hardwick2020_DTC}.
The power-limiting effects of point absorbers (Section~\ref{ss:absorption}) will likely be mitigated by replacing the affected test masses.

Various upgrades to improve the observed squeezing level are planned (Section~\ref{ss:squeezer}).
These include new output Faraday isolators to reduce losses, higher green power to increase the injected squeezing level, and deformable optics to improve mode matching between the squeezer and interferometer.

Additional observatory upgrades are planned to prepare the site for A+, the detector configuration that will be used after O4 \cite{Miller2015,Barsotti2018}.
As the optical power and squeezing level increase, quantum radiation pressure noise (Section~\ref{ss:quantum}) will worsen low-frequency sensitivity.
To mitigate this effect, a \SI{300}{m} filter cavity for injection of frequency-dependent squeezing \cite{EvansFilterCav, mcculler2020frequencydependent} will be installed prior to O4.

Work continues to understand and mitigate known noise sources.
Additional stray light baffles installed where vibrational coupling has been observed should mitigate the effect of scattered light (Section~\ref{ss:scatterNoise}).
Improved modeling of angular motion coupling to DARM and how this may be mitigated is underway.
The use of universal (as opposed to local) control on seismic isolation platforms at the vertex, similar to the scheme used to ride out earthquakes (Section~\ref{ss:seismicupgrade}), is expected to improve the nonlinear noise coupling during times of large ground motion and also improve interferometer duty cycle by limiting saturations.
Machine learning techniques are being developed that allow offline removal of nonlinear noise contributions \cite{vajente2019machinelearning}.

The challenge to discover and mitigate the sources of noise below \SI{100}{Hz} will also be critical for Advanced LIGO to achieve design sensitivity in O4.
This achievement's potential reward is another 40\% increase in the astrophysical range of the detectors and commensurate tripling of the expected number of detections.
O3 was the most successful search for astrophysical gravitational-wave sources in history;
O4 promises even greater knowledge of the furthest reaches of the universe.
 
\section{Acknowledgements}
The authors gratefully acknowledge the support of the United States
National Science Foundation (NSF) for the construction and operation of the
LIGO Laboratory and Advanced LIGO as well as the Science and Technology Facilities Council (STFC) of the
United Kingdom, and the Max-Planck-Society (MPS) for support of the construction of Advanced LIGO.
Additional support for Advanced LIGO was provided by the Australian Research Council.
The authors acknowledge the LIGO Scientific Collaboration Fellows program for additional support.

LIGO was constructed by the California Institute of Technology and Massachusetts Institute of Technology with funding from the National Science Foundation, and operates under cooperative agreement PHY-1764464. Advanced LIGO was built under award PHY-0823459. This paper carries LIGO Document Number LIGO-P2000122. 
\appendix
\section{Table of O3 Parameters}
\label{s:table}

\begin{table*}[t]
		\centering
    \begin{tabular}{ l c c c c }

      Parameter & Symbol & LHO Value & LLO Value & Units  \\ \hline

      Squeezing Levels & $\mathrm{dB}_\mathrm{SQZ}$ & 2.0 & 2.7 & dB \\

      1st Modulation Sideband Frequency & $f_9$ & 9.100230 & 9.099055 & MHz \\ 2nd Modulation Sideband Frequency & $f_{45}$ & 45.501150 & 45.496925 & MHz \\ 3rd Modulation Sideband Frequency & $f_{118}$ & 118.302990 & 118.287715 & MHz \\ 

      1st Modulation Depth & $\Gamma_9$     & 0.135 & 0.14 & rads \\ 2nd Modulation Depth & $\Gamma_{45}$  & 0.177  & 0.16 & rads \\
      3rd Modulation Depth & $\Gamma_{118}$ & 0.012  & 0.019 & rads \\

      ETMX Transmission & $T_\mathrm{ETMX}$ & 3.9 & 4.0 & ppm \\
      ETMY Transmission & $T_\mathrm{ETMY}$ & 3.8 & 3.9 & ppm \\
      ITMX Transmission & $T_\mathrm{ITMX}$ & 1.50 & 1.48 & \% \\
      ITMY Transmission & $T_\mathrm{ITMY}$ & 1.42 & 1.48 & \% \\
      PRM Transmission & $T_\mathrm{PRM}$ & 3.1 & 3.1 & \% \\
      SRM Transmission & $T_\mathrm{SRM}$ & 32.34 & 32.40 & \% \\

      Arm Length & $L$ & 3994.5 & 3994.5 & m \\
      Power-Recycling Cavity Length & $l_\mathrm{P}$ & 57.7 & 57.7 & m \\
      Signal-Recycling Cavity Length & $l_\mathrm{S}$ & 56.0 & 56.0 & m \\
      Schnupp Asymmetry & $l_\mathrm{schnupp} = l_\mathrm{x} - l_\mathrm{y}$ & 0.08 & 0.08 & m \\

      Arm Free Spectral Range & $f_\mathrm{FSR}$ & 37.5 & 37.5 & kHz \\

      X Arm Cavity Pole & $f_{\mathrm{X}}$ & 45.1 & 44.5 & Hz \\
      Y Arm Cavity Pole & $f_{\mathrm{Y}}$ & 42.7 & 44.5 & Hz \\

      CARM Cavity Pole & $f_{\mathrm{CARM}}$ & 0.6 & 0.4 & Hz \\
      DARM Cavity Pole & $f_{\mathrm{DARM}}$ & 411 & 455 & Hz \\

      ETMX Green Transmission & $T^g_\mathrm{ETMX}$ & 7.9 & 4.8 & \% \\
      ETMY Green Transmission & $T^g_\mathrm{ETMY}$ & 7.9 & 5.0 & \% \\
      ITMX Green Transmission & $T^g_\mathrm{ITMX}$ & 0.96 & 0.95 & \% \\
      ITMY Green Transmission & $T^g_\mathrm{ITMY}$ & 1.10 & 1.11 & \% \\

      X Arm Green Cavity Pole & $f^g_{\mathrm{X}}$ & 274.6 & 175.4 & Hz \\
      Y Arm Green Cavity Pole & $f^g_{\mathrm{Y}}$ & 278.8 & 186.5 & Hz \\

      Input Mode Cleaner Modulation Frequency & $f_{24}$ & 24.1 & 24.1 & MHz \\
      Input Mode Cleaner Modulation Depth & $\Gamma_{24}$ & 13 & 16 & mrads \\

      Input Mode Cleaner Round Trip Length & $L_\mathrm{IMC}$ & 32.9434 & 32.9465 & m \\

      Input Mode Cleaner Cavity Pole & $f_\mathrm{IMC}$ & 8625.2 & 8919.4 & Hz \\

    \end{tabular}

		\caption{
    Summary optical and physical parameters of the Advanced LIGO interferometers during O3.
    }
		\label{table:aLIGOParams}
\end{table*}

\bibliography{references}

\end{document}